\documentclass{aa}  

\usepackage{graphicx}
\usepackage{multirow}
\usepackage{hyperref}
\usepackage{txfonts}
\graphicspath{{Figures/}{../Figures/}}
\begin{document} 
    
   \title{Geometry versus growth}

   \subtitle{Internal consistency of the flat $\Lambda$CDM model with KiDS-1000}

   \author{J. Ruiz-Zapatero\inst{1,2} \and
          Benjamin St\"olzner \inst{2} \and
          Benjamin Joachimi \inst{2} \and
          Marika Asgari \inst{3} \and
          Maciej Bilicki \inst{4} \and
          Andrej Dvornik \inst{5} \and
          Benjamin Giblin \inst{3} \and
          Catherine Heymans\inst{3,5} \and
          Hendrik Hildebrandt \inst{5} \and
          Arun Kannawadi \inst{6} \and
          Konrad Kuijken \inst{7} \and
          Tilman Tr\"oster \inst{3} \and
          Jan Luca van den Busch \inst{5} \and
          Angus H. Wright \inst{5}}

   \institute{University of Oxford, Denys Wilkinson Building, Keble Road, Oxford, OX1 3RH, UK\\
              \email{jaime.ruiz-zapatero@physics.ox.ac.uk}, 
              \and
              Department of Physics and Astronomy, University College London, Gower Street, London WC1E 6BT, UK 
              \and
              Institute for Astronomy, University of Edinburgh, Royal Observatory, Blackford Hill,  Edinburgh, EH9 3HJ, UK
              \and
              Center for Theoretical Physics, Polish Academy of Sciences, al. Lotników 32/46, 02-668, Warsaw, Poland
              \and
              Ruhr University Bochum, Faculty of Physics and Astronomy, Astronomical Institute (AIRUB), German Centre for Cosmological Lensing, 44780 Bochum, Germany
              \and
              Department of Astrophysical Sciences, Princeton University, 4 Ivy Lane, Princeton, NJ 08544, USA
              \and
              Leiden Observatory, Leiden University, P.O.Box 9513, 2300RA Leiden, The Netherlands}

   \date{Received --; accepted ---}

 
  \abstract
   {We carry out a multi-probe self-consistency test of the flat $\Lambda$CDM model with the aim of exploring potential causes of the reported tensions between high- and low-redshift cosmological observations.
We divide the model into two theory regimes determined by the smooth background (geometry) and the evolution of matter density fluctuations (growth), each governed by an independent set of Lambda Cold Dark Matter ($\Lambda$CDM) cosmological parameters.
This extended model is constrained by a combination of weak gravitational lensing measurements from the Kilo-Degree Survey, galaxy clustering signatures extracted from Sloan Digital Sky Survey campaigns and the Six-Degree Field Galaxy Survey, and the angular baryon acoustic scale and the primordial scalar fluctuation power spectrum measured in \textit{Planck} cosmic microwave background (CMB) data.
For both the weak lensing data set individually and the combined probes, we find strong consistency between the geometry and growth parameters, as well as with the posterior of standard $\Lambda$CDM analysis. In the non-split analysis, for which one single set of parameters was used, tension in the amplitude of matter density fluctuations as measured by the parameter $S_8$ persists at around $3\sigma$, with a $1.5\,\%$ constraint of $S_8 = 0.776_{-0.008}^{+0.016}$ for the combined probes.
We also observe a less significant preference (at least $2\sigma$) for higher values of the Hubble constant, $H_0 = 70.5^{+0.7}_{-1.5}\,{\rm km\, s^{-1} Mpc^{-1}}$, as well as for lower values of the total matter density parameter $\Omega_{\rm{m}} = 0.289^{+0.007}_{-0.005}$ compared to the full \textit{Planck} analysis. Including the subset of the CMB information in the probe combination enhances these differences rather than alleviate them, which we link to the discrepancy between low and high multipoles in \textit{Planck} data.
Our geometry versus growth analysis does not yet yield clear signs regarding whether the origin of the discrepancies lies in $\Lambda$CDM structure growth or expansion history but holds promise as an insightful test for forthcoming, more powerful data.}


   \maketitle
%

\section{Introduction}

The dramatic increase in precision experienced by cosmology over the last 30 years has recently led to the discovery of potential inconsistencies in our cosmological model that previously might have been obscured by statistical errors. One of the most relevant of these inconsistencies is the apparent discrepancy between a subset of the free parameters of the concordance Lambda Cold Dark Matter ($\Lambda$CDM) model \citep{LCDM, Ofer} measured at early and late times of the Universe. The most notable manifestation of this tension is the $4.2 \sigma$ \citep{riess_42} difference in the value of the Hubble constant, $H_0$, between distance ladder estimates and the early-Universe cosmic microwave background (CMB) probe \textit{Planck} \citep{Planck}. Moreover, late-Universe probes of the large-scale structure, such as weak gravitational lensing (WL) and galaxy clustering, prefer a lower amplitude of the growth of structure than \textit{Planck}, with tension up to the $3.2\sigma$ level in the parameter $S_{\rm{8}} = \sigma_{\rm{8}} \sqrt{\frac{\Omega_{\rm{m}}}{0.3}}$, where $\sigma_{\rm{8}}$ is the standard deviation of matter density fluctuations in spheres of radius $8\,h^{-1}{\rm Mpc}$ today and $\Omega_{\rm{m}}$ is the matter density parameter \citep{ CFHTLenS, Joudaki, Loureiro2019, K1K, KV+BOSS,  Kobayashi2020}. 

If confirmed, the reasons behind these discrepancies could be twofold. On the one hand, if we assume our cosmological theory to be correct, the origin of the tension must lie in the data or their analysis of either set of probes. On the other hand, if we decide to trust the current measurements, the tension would suggest a shortcoming of the theory used to analyse the data, such as new physics beyond the standard model \citep[e.g.][]{HTension, tension_new_physics, Decaying_S8, Dianotti21}. Finally, one must also consider the possibility that both measurements and theory are correct and that the disagreement is solely due to statistical variance between the two sets of measurements.

Numerous efforts have been made to identify possible errors in the respective data analysis processes of each set of probes. Concerning the local measurements of the Hubble parameter, examples of these efforts can be found in the most recent review of the distance ladder \citep{riess_42}. Large-scale structure surveys have also carried out extensive systematics checks and internal consistency tests; see \citet{DESY1_cos, DES_Planck_tension} and \citet{DES_consistency} for the Dark Energy Survey, \citet{ Consistency, wright20, Busch20, giblin21, K1K} and \citet{ hildebrandt20} for the Kilo-Degree Survey (KiDS),  and \citet{eBOSS_sys} and  \citet{boss_dr12_sys} for the Baryon Oscillation Spectroscopic Survey (BOSS) and the Extented Baryon Oscillation Spectroscopic Survey (eBOSS) respectively. As for the CMB probes, \citet{Efstathiou} undertook a thorough revision of the CamSpec Likelihood used to analyse \textit{Planck} data in search of potential sytematics. Moreover, the initiative 'Beyond \textit{Planck}' is currently conducting a thorough revision of the \textit{Planck} and WMAP methodology \citep{BeyondPlanck}.

Recently, alternative calibrations of the cosmic distance ladder have shed doubts on the robustness of the Hubble tension \citep[e.g.][]{Wendy19, Wendy21}. Moreover, alternative methods of measuring the Hubble parameter locally seem to be in agreement with the CMB measurement rather than with the cosmic distance ladder \citep[e.g.][]{DESY1H0, Hall21}. However, no known systematic errors have convincingly explained the anomalously large value of $H_0$ of the cosmic distance ladder yet \citep[e.g.][]{TRGB_H, masers_H}. Similarly, the $S_{\rm{8}}$ tension has proven not to be relieved by combining data of independent late-Universe probes \citep[e.g.][]{DES, KV+BOSS, K1K+BOSS}. 

In this work we develop and apply methodology to test the self-consistency of the spatially flat $\Lambda$CDM model with the goal of providing guidance as to which regime of the theory could drive the observed tension. To do so, we differentiate between two theory regimes within the cosmological model: geometry; refereeing to the parts of the model related to the dynamics of the space-time and growth, concerning the formation of matter anisotropies on top of the dynamics of space-time. The key distinguishing factor between the two regimes is that geometry solely regards uniform background quantities while growth considers perturbations on this background. If the $\Lambda$CDM model is self-consistent, the same set of cosmological parameter values should correctly reproduce both of its theory regimes. We test this assumption by letting each theory regime be governed by an independent set of $\Lambda$CDM parameters. Then, we employ a range of high- and low-redshift cosmological probes to obtain constraints for each regime of the theory. Any discrepancies between the pairs of geometry and growth parameters would hint at the origin of potential failures in our current cosmological model. Moreover, the fact that this methodology studies background and perturbations independently allows us to more easily pinpoint the character and origin of the new physics behind the potential discrepancy.

Similar studies were conducted prior to this publication that also explore the effects of splitting certain parameters of a given cosmological model as a means to investigating its self-consistency. As early as 2004, \citet{chu_04} studied the consistency of the constraints coming from data on the ionisation history and on the pressure profile of the pre-recombination fluid in the context of flat $\Lambda$CDM  by duplicating the baryon density parameter, $\Omega_{\rm{b}}$, and allowing each instance to be constrained by one of the phenomena. In \citet{3omegas} a similar exercise was performed on the total matter density parameter, $\Omega_{\rm{m}}$, by having three instances, each controlled by a different physical observable in type Ia supernova (SN Ia) data. More recently, \citet{Huterer} undertook a very similar analysis to the one presented in this work, with a similar selection and treatment of observables but limited to splitting  $\Omega_{\rm{m}}$. Beyond $\Lambda$CDM,  \citet{bernal_16} and \citet{wang_07} undertook multi-probe studies of the consistency of the variable equation of state, cold dark matter ($w$CDM) model by splitting the equation-of-state parameter $w$, and the density parameter, $\Omega_{\rm{DE}}$, of dark energy into their geometry and growth contributions. 

During the final stages of our study, \citet{DES_geo_gro} published a closely analogous investigation of geometry--growth consistency using several probes from the Dark Energy Survey in combination with external data. Our work is complementary in that it employs a different mix of data sources: WL data from KiDS-1000 \citep{K1K}, galaxy clustering data from BOSS  \citep{BOSS} and the 6 Degree Field Galaxy  Survey (6dFGS; \citealp{6dF_BAO}), Lyman-$\alpha$ and quasar clustering information from the Extended Baryon Oscillation Spectroscopic Survey (eBOSS; \citealp{ebossdata1,ebossdata2}), and subsets of the \textit{Planck} CMB temperature and polarisation anisotropy correlations. Moreover, our split is more comprehensive in that we duplicate the full flat-$\Lambda$CDM parameter space, whereas  \citet{Huterer} and \citet{DES_geo_gro} limited themselves to a split in the dark matter density parameter only.

This work is structured as follows: In Section~\ref{Methodology} we present the methodology of this work regarding the distinction between geometry and growth in the $\Lambda$CDM model and the resulting modelling of the observables of interest. In Section~\ref{Data Sets} we describe the data sets that we analyse. We discuss the likelihood analysis in Section~\ref{Likelihood analysis} and present results in Section~\ref{Results}. In Section~\ref{Conclusions} we summarise our findings and conclude.

\section{Methodology} 
\label{Methodology}

\subsection{Distinguishing geometry and growth} \label{Distinguishing geometry and growth}

We adopted a classification of the theory of the $\Lambda$CDM model based on whether it keeps, or departs from a uniform background. Hence, any information regarding the formation of structure in the model can be traced to where it departs from a smooth background. Thus, we distinguished two theoretical regimes: on the one hand, a uniform background that describes the curvature and expansion history of the Universe, and on the other hand, a theory of perturbations that build up from the background and create the structures in the Universe\footnote{As in standard structure formation theory, we assumed here that structure growth does not feed back significantly on the expansion history; see \citet{Backreaction} for a detailed discussion about this backreaction.}. We refer to these two regimes of the theory as geometry and growth, respectively.

In mathematical terms the departure from, or conservation of, such uniformity can be fully encapsulated in the choice of metric since its shape is fully determined by the assumptions that we make about the universe that we aim to describe. Furthermore, the fundamental role that the metric plays in Einstein gravity means that these assumptions propagate into the entirety of the theoretical predictions of the model. Of these predictions, we particularly focus on two: the field equations, that describe the relationship between matter and gravity, and the equations of motion of the stress-energy tensor fields, that describe how the content of the Universe evolves over time. The FLRW line element, that represents a uniform background, directly leads to the Friedmann field equations. Moreover, the Friedmann field equations can be combined into a conservation law for the field that describes its evolution. The perturbation of this uniform background metric ultimately leads to the Jeans equation, describing the evolution of matter density perturbations over cosmic time \citep[e.g.][]{baker_ferreira}.

This dualistic understanding of the theory is extremely useful when calculating predictions of astrophysical phenomena. This is due to the fact that some phenomena can be accurately modelled purely within a particular regime. Thus, we can equivalently distinguish between geometry and growth observables as well as  geometry and growth theory regimes. Furthermore, the modelling of more 'complex' observables might need a combination of background and perturbation physics. 
 
Nonetheless, if the $\Lambda$CDM model is to be self-consistent, both regimes of the theory must be governed by the same set of parameters, and therefore, parameter constraints based on geometry observables should be consistent with those from growth ones. Similarly, observables whose modelling makes use of both theory regimes should exhibit internal consistency between the preferred parametrisations of their different calculation stages. 

In this work we tested the self-consistency of the $\Lambda$CDM model by assigning to each theory regime its own set of cosmological parameters and simultaneously parametrising them using a set of geometry and growth observables. This allowed us to observe the parametrisation preferences of each regime such that potential discrepancies can be identified. 
From here on we denote the sets of parameters governing the geometry and growth regimes of the theory as $\boldsymbol{p}^{\rm{geom}}$ and $\boldsymbol{p}^{\rm{grow}}$, respectively. In this analysis we sample over the following parameters: $S_8= \sigma_{\rm{8}} \sqrt{\Omega_{\rm{m}}/0.3}$; where $\sigma_{\rm{8}}$ is the root-mean-square of matter density fluctuations in spheres of $8\,h^{-1}{\rm Mpc}$ and $\Omega_{\rm{m}}$ is the total matter density parameter, the reduced Hubble parameter $h = H_{\rm{0}}/(100\,{\rm km\, s^{-1} Mpc^{-1}})$, the reduced cosmological cold dark matter density $\omega_{\rm{cdm}} =  \Omega_{\rm{cdm}} h^2$, the reduced cosmological baryonic density $\omega_{\rm{b}} =  \Omega_{\rm{b}} h^2 $  and the spectral index $n_{\rm{s}}$.

Simply splitting the parameter space into two subsets does not ensure independent constraints for the two theory regimes. This is because one can fully derive the growth aspects from the geometry parameters and vice versa, via the Jeans equation, in cosmological models in which gravitational collapse is quasistatic up to the linear level of the theory \citep{silvestri, baker_ferreira}. An explicit demonstration of this duality in GR+$\Lambda $CDM can be found in \citet{starobinsky}. Moreover, even if the set of geometry cosmological parameters cannot constrain parameters that explicitly relate to matter anisotropies (e.g. $S_{\rm{8}}$), fixing them would misrepresent this lack of constraining power. An analogous argument can be made for growth parameters that primarily affect the expansion history, such as $h$. We therefore assigned the same, uninformative priors to both parameter sets and vary the full set in each case.

There are alternative interpretations of the categories of geometry and growth and that the choice made in this work is not necessarily unique. For example, in \citet{DES_geo_gro} growth is limited to describe only the late growth of structure based on the made argument that geometry and growth are analytically linked at the level of linear theory. These approaches are not contradictory but the different analysis choices prevent direct comparisons between the resulting constraints.

For the purposes of this work we limited ourselves to a spatially flat $\Lambda$CDM model. In agreement with \citet{KV}, we also assumed two massless neutrino species and a third massive species with mass $m_{\rm{ncdm}} = 0.06 $ eV$/c^2$,  as well as a  temperature of the non-cold relic $T_{\rm{ncdm}} = 0.71611$ K.

\subsection{Cosmological observables} \label{Cosmological Observables}

In this section we introduce the physical observables used in this work, including a discussion of their geometry versus growth treatment; see Table~\ref{tab:theory} for an overview.

\begin{table} 
\caption{Geometry and growth classification of the observables used in our analysis.}
\label{tab:theory}
\centering
\begin{tabular}{ p{0.4\columnwidth} p{0.4\columnwidth}  }
 \hline
  \hline
 Geometry & Growth \\
 \hline
BAO angular scale  & RSD growth rate\\ 
WL lensing efficiencies  & WL matter power spectrum \\
CMB first acoustic peak position &  CMB primordial power spectrum \\
\hline
\end{tabular}
\tablefoot{The first row lists features extracted from clustering signals (BAO: baryon acoustic oscillations; RSD: redshift space distortions), the second row weak lensing (WL) signal contributions, and the third row quantities inferred from cosmic microwave background (CMB) measurements.}
\end{table}

\subsubsection{Weak lensing} \label{Weak lensing}

In general relativity mass induces in its surrounding space-time a non-Euclidean geometry. In agreement with Fermat's principle, photons seek the path of least action through the curved space. These paths correspond to geodesic lines that are not straight in a Euclidean sense, thus causing light rays to appear bent. This deflection of light rays leads to similar observational effects as those of optical lenses, focusing and distorting the images we observe of celestial bodies that lie behind the lens. The effect of such lenses is usually quite weak, such that it can only be detected via statistical methods from large galaxy samples. This is referred to as weak lensing (WL). 

Weak lensing measurements are an excellent tool to inspect the interplay between the different regimes of $\Lambda$CDM because it is sensitive to both geometry and growth. This is because the strength of the light deflection depends on the distances between the light source, the lens, and the observer, as well as the spatial distribution of the lenses and their density contrast with respect to the mean matter distribution. Formally, this duality can be observed in the key quantity of weak lensing, the lensing convergence $\kappa$ \citep[e.g.][]{Bartelmann2001},
\begin{equation}
\label{eq:kappa}
        \kappa^i(\boldsymbol{\theta}) =  \frac{3 H_0^2 \Omega_{\rm m}}{2c^2}  \int^{\chi_{\rm lim}}_0  \frac{{\rm d}\chi  \, \chi}{a(\chi)} g^i(\chi) \delta(\chi \boldsymbol{\theta}, \chi) \;,
\end{equation}
where we assume a spatially flat universe. Here, $\chi$ is the radial comoving distance, $\chi_{\rm{lim}}$ is the radial comoving distance to the horizon, $a$ is the scale factor, $c$ is the speed of light in the vacuum, and $\boldsymbol{\theta}$ is a two-dimensional vector that represents angular position on the sky. We also introduce the matter density contrast $\delta$ measured at a position and epoch determined by the comoving distance and the angular position. The lensing efficiency $g^i(\chi)$ is defined as
\begin{equation} 
\label{eq:lens_eff}
    g^i(\chi) = \int^{\chi_{\rm lim}}_\chi {\rm d}\chi' n^i(\chi')\, \frac{\chi'-\chi}{\chi'}\;,
\end{equation}
where $n^i(\chi')$ is the comoving distance distribution of source galaxies on which the lensing effect is measured, assigned to a tomographic redshift bin labelled by $i$. In practice, we measure the source distribution of redshifts, $n^i(z)=n^i(\chi) \, {\rm d} \chi/{\rm d}z$.

Since the effects of weak lensing are not strong enough to be directly observable, it is only possible to appreciate them statistically. The standard baseline two-point statistic to model weak lensing signals is the Fourier transform of the two-point correlation of the convergence  across redshift bins $i$ and $j$, the convergence power spectrum 
\begin{equation}
\label{eq:kappakappa}
        C_\kappa^{ij}(\ell) = \frac{9}{4}\Omega^2_{\rm m} \left( \frac{H_0}{c} \right)^4 \int^{\chi_{\rm lim}}_0 \!\!\! {\rm d} \chi \frac{g^i(\chi) g^j(\chi) }{a^2(\chi)} P_{\rm{\delta}}\left( \frac{\ell + 1/2}{\chi},\chi \right) \;,
\end{equation}
where $P_{\delta}(k,\chi)$ is the (non-linear) matter power spectrum evaluated at wavenumber $k$ and an epoch marked by $\chi$. In Eq.~(\ref{eq:kappakappa}) the Limber approximation was applied \citep{Limber}.

Angular power spectra are not directly measurable from a restricted survey footprint. We employed band powers derived from angular correlation function measurements; see \citet{K1K_methodology,bps_b,bps_a} for a detailed description. Band powers can be modelled as linear functionals of the angular convergence power spectra,
\begin{equation}
\label{eq:bps}
{\cal C}_{{\rm E},l}^{ij} =  \frac{1}{2 {\cal N}_l} \int_0^\infty {\rm d}
\ell \ell \; W^l_{\rm EE}(\ell)\; C^{ij}_{\kappa}(\ell)\;,
\end{equation}
for a band indexed by $l$, where the filter function $W^l_{\rm EE}$ is given by equation 26 of \citet{K1K_methodology}. The normalisation, $\mathcal{N}_l$, is defined such that the band powers
trace $\ell^2C_\kappa(\ell)$ at the logarithmic centre of the bin,
$\mathcal{N}_l = \ln(\ell_{{\rm up},l}) - \ln(\ell_{{\rm lo},l})$
with $\ell_{{\rm up},l}$ and $\ell_{{\rm lo},l}$ defining the edges of the top-hat band selection function for the bin indexed by $l$. Here, we have assumed that the model does not predict any B-modes, and we will only use the E-mode band powers in our analysis. This assumption is based on the work of \citet{giblin21} who showed that there is no significant detection of B-modes in KiDS-1000.

It is possible to classify the quantities that enter the convergence power spectrum as geometry- or growth-related. On the one hand, the cosmological dependence of the matter power spectrum $P_{\delta}$ is growth-related. On the other hand, the lensing efficiency  solely concerns angular diameter distances (or comoving distances when assuming a flat universe), that are purely geometry-related. A more ambiguous decision is whether to consider the prefactor in Eq.~(\ref{eq:kappakappa}) geometry or growth. This is due to the fact that, while these terms originate from the Poisson equation that relates the gravitational potential to its source, $\delta$, in the context of the calculation of the observable they effectively act as a geometrical relationship that does not necessarily regard anisotropies. In agreement with the criteria of recent work \citep{DES_geo_gro}, we consider these prefactors as geometry.

\subsubsection{Baryon acoustic oscillations} \label{Baryonic acoustic oscillations}

Baryon acoustic oscillations (BAOs) cause enhanced matter overdensities  at a characteristic physical separation scale that corresponds to the size of the sound horizon at the end of the drag epoch, $r_{\rm{s}}(z_{\rm d})$ \citep{Peebles,CMB}. The sound horizon is defined as the distance a pressure wave can travel from its time of emission in the very early Universe up to a given redshift. This can be expressed as 
\begin{equation} \label{eq:rsound}
    r_{\rm{s}}(z)= \int^{\infty}_z \frac{c_{\rm{s}}  \, {\rm d}z'}{H(z')} \;,
\end{equation}
where $c_{\rm s}$ denotes the speed of sound, and where $H(z)$ is the expansion rate at redshift $z$. The end of the drag epoch is defined as the time when photon pressure can no longer prevent gravitational instability in baryons around $z \sim 1020$ \citep{z_d}. The redshift of this time is typically estimated using the following fitting formula:
\begin{equation} \label{eq:zsound}
    z_{\rm{d}}= \frac{1291(\omega_{\rm{m}})^{0.251}}{1 + 0.659(\omega_{\rm{m}})^{0.828}} [1+ b_{\rm{1}} (\omega_{\rm{b}})^{b_{\rm{2}}}
    ]  \;,
\end{equation}
where $b_{\rm{1}} = 0.313 (\omega_{\rm{m}})^{-0.419}[1+ 0.607(\omega_{\rm{m}})^{0.674}]$, $b_{\rm{2}} = 0.238 (\omega_{\rm{m}})^{0.223}$ and $\omega_{\rm{m}} =  \omega_{\rm{cdm}} + \omega_{\rm{b}}$ is the reduced cosmological matter density (see \citet{z_d}).

Since the overdensity manifests anisotropically in redshift space as a three-dimensional shell, 
it is possible to measure the shift in the BAO peak position with respect to its
position in a fiducial cosmology, parallel and perpendicular to the line of sight. Perpendicular to the line of sight, the BAO feature informs the trigonometric relationship
\begin{equation}\label{eq:acoustic}
    \theta \approx \frac{r_{\rm{s}}(z)}{D_{\rm{M}}(z)} \;,
\end{equation}
where $\theta$ is the angle under which the scale of the sound horizon is observed,  and $D_{\rm{M}}(z)$ is the comoving angular diameter distance to redshift $z$, which in a flat universe is identical to the comoving radial distance $\chi(z)$. Parallel to the line of sight, the BAO feature allows us to measure the expansion history of the Universe. The BAO measurements along the line of sight can be use to constrain the relationship $H(z) r_{\rm{s}}(z_{\rm{d}})$.

We cast our BAO measurements in the form of the dimensionless ratios \citep{BOSS, eBOSS16a, ebossdata1, ebossdata2}:
 \begin{equation} \label{eq:alphas}
     \alpha_\parallel = \frac{[D_{\rm{H}} / r_{\rm{s}}(z_{\rm{d}}) ]}{[D_{\rm{H}} / r_{\rm{s}}(z_{\rm{d}})]_{\rm{fid}}}; ~~~
    \alpha_\perp = \frac{[\chi / r_{\rm{s}}(z_{\rm{d}}) ]}{[\chi /r_{\rm{s}}(z_{\rm{d}})]_{\rm{fid}}} ,
\end{equation}
to describe shifts perpendicular and parallel to the line of sight, where  $D_{\rm{H}}(z) = c/H(z)$ is the Hubble distance and the label `fid' denotes a quantity evaluated at the aforementioned fiducial cosmology. Nonetheless, the quantities chosen to represent these parallel and perpendicular signatures of the BAO feature can vary from probe to probe. 

It is also common to combine the perpendicular and parallel information in the  volume-averaged angular diameter distance defined as
\begin{equation} \label{eqn:dv}
    D_{\rm{v}}(z) = \left( \frac{c z}{H(z)} \chi^2(z) \right)^{1/3} \;,
\end{equation}
or in the anisotropy parameter known as the Alcock-Paczynski parameter ; $F_{\rm{AP}}(z)$  \citep{AP_effect}, defined as:
\begin{equation} \label{eqn:F_AP}
    F_{\rm{AP}}(z) =  \chi(z) H(z) /c \;.
\end{equation}

Although BAOs are a consequence of structure growth, their signature in the matter distribution can be translated into distance relationships that can be calculated without making use of density perturbations. This can be seen in the fact that, while the speed of propagation of BAOs, and thus the sound horizon, is a function of the matter content of the Universe, it can be completely calculated by assuming this content to be  uniform. Therefore, measurements of the position of the BAO peak in the CMB or in the large-scale structure can be classified as pure geometry phenomena.

\subsubsection{Redshift space distortions} \label{Redshift space distortions}

Redshift space distortions (RSDs) are modifications to the clustering statistics of a given ensemble of objects due to their peculiar velocities along the line of sight on top of the recession velocity due to cosmological expansion \citep{kaiser}. RSDs are determined by peculiar motion in the gravitational potential of the surrounding matter distribution and thus governed by the Poisson equation, which makes them a pure \textit{growth} effect. For galaxies on large, linear scales RSDs are dominated by the infall towards overdense structure, known as the Kaiser effect \citep{RSD}. 

Clustering two-point statistics as a function of transverse and line-of-sight separation can be used to extract RSDs via the quantity \citep[e.g.][]{rsd_florian}
\begin{equation}
\label{eqn:rsd}
    f \sigma_8 \equiv  \frac{{\rm d} \ln \delta}{{\rm d} \ln a} \sigma_8 \;,
\end{equation}
where $f$ is the growth rate and $\delta$ is again the matter density contrast. Both $f$ and $\sigma_8$ are evaluated at the effective redshift of the measurement, as opposed to the use of $\sigma_8$ in weak lensing that is always interpreted at $z=0$.

\subsubsection{Early-Universe geometry and growth parameters} \label{Primordial anisotropies parameters}

The CMB is the richest source of cosmological information to date and particularly valuable as a complement to low-redshift probes of the large-scale structure. While a full re-analysis of \textit{Planck} data is beyond the scope of this work, we would still like to make use of readily accessible CMB information that can be allocated cleanly to either the  geometry or growth regimes. The primary CMB anisotropies can be described by a set of five cosmological parameters \citep{CMB_params, CMB_nat_params}: the reduced baryon density parameter $\omega_{\rm{b}}$, the reduced cold dark matter density parameter $\omega_{\rm{cdm}}$, the amplitude and spectral index of primordial scalar fluctuations, $A_{\rm{s}}$ and $n_{\rm{s}}$ respectively, and the angle subtended by the sound horizon at end of recombination epoch $\theta^*$ (cf. Eq.~\ref{eq:acoustic}, evaluated at redshift $z^* \sim 1100$ as opposed to $z_{\rm{d}} \sim 1020$).
A detailed discussion of the role of each of these parameters in determining the physical properties of the CMB can be found in \citet{CMB_params_detail_a}.

The CMB information contained in the parameters $n_{\rm{s}}$ and $A_{\rm{s}}$ is purely growth-related since they capture the primordial scalar fluctuation power spectrum 
  \begin{equation} \label{eqn:Harr_Zel}
      P_{\delta, {\rm prim}}(k) = A_{\rm{s}} \left(\frac{k}{k_{\rm{0}}}\right)^{n_{\rm{s}}} \;,
 \end{equation}
where $k_{\rm{0}}$ is the pivot scale of the power spectrum here set to the value of $0.05$ Mpc$^{-1}$ for consistency with \textit{Planck} \citep{Planck}. In contrast, $\theta^*$ is clearly a parameter solely concerned with the background cosmology as it relates the distance to the surface of last scattering to the size of the sound horizon. 

The information contained in $\omega_{\rm{b}}$ and $\omega_{\rm{cdm}}$ cannot not readily categorised. On the one hand, they are informed by the relative height of the CMB peaks, related to density fluctuations, but also by the expansion history and the value of the speed of sound in the baryon-photon fluid, all describable in a smooth universe. Therefore we decided not to include CMB constraints on $\omega_{\rm{b}}$ and $\omega_{\rm{cdm}}$ in our analysis, but do make use of the marginalised posteriors of the combination $ \{ n_{\rm{s}}, A_{\rm{s}}, \theta^* \}$, taking their correlation into account. We note that similar analyses \citep{DES_geo_gro} have employed the CMB shift parameter $\mathcal{R}$, a measure of the change of location of the first power spectrum peak with respect to a fiducial cosmology \citep{CMB_shift}. While the $\mathcal{R}$ parameter is a good alternative to constrain geometry, in this work we decided to use $\theta^*$ as it is a direct analogue of quantities available for the large-scale structure BAO measurements. The $n_{\rm{s}}$ constraint can be directly matched to the sampled $n_{\rm{s}}$ values in our analysis to constrain growth. Similarly, $A_{\rm{s}}$ can be inferred from the sampled $S_{\rm{8}}$ value and matched to the CMB constraint to constrain growth. On the other hand, we treat $\theta^*$ as another evaluation of the distance relationship in Eq.~(\ref{eq:acoustic}) and use it to constrain the geometry regime.

\section{Data Sets} \label{Data Sets}

\begin{table*} 
\centering
\caption{Data sets used in our analysis, and their properties.}
\label{tab:data}
\begin{tabular}{ p{3cm}p{4.5cm}p{2.5cm}p{3.5cm}p{1cm}  }
 \hline
 \hline
 Probe & Data set & Redshifts & Observable(s) & Data Points\\
 \hline
Weak lensing   & KiDS-1000 \citep{kuijken19, hildebrandt20, giblin21} & 0.1 - 1.2 (photometric redshift) & Tomographic band power spectra ${\cal C}_{{\rm E},l}^{ij}$ (see Eq.~\ref{eq:bps})
& 120  \\
  \hline
Clustering - BAO  &
BOSS DR12 \citep{BOSS} & 0.38 - 0.61 & $[H(z)]_{\rm{fid}}/(\alpha_\parallel)$;  $\alpha_\perp [\chi(z)]_{\rm{fid}}$ (see Eq.~\ref{eq:alphas_B})
& 6\\
 &  6dfGS \citep{6df_data} & 0.106 & $r_{\rm{s}}(z_{\rm{d}})/D_{\rm{v}}$  & 1 \\

Clustering - RSD  &  BOSS DR12 \citep{BOSS} & 0.38 - 0.61 & $f\!\sigma_{\rm{8}}$ & 3 \\ 
\hline
Ly$\alpha$ - BAO   & eBOSS DR14 \citep{ebossdata1, ebossdata2}     & 2.34  &  $\alpha_\parallel$;  $\alpha_\perp$  & 6360  \\
\hline
CMB - BAO   &  \textit{Planck} 2018 TT, TE, EE + Lowl + lowE \citep{Planck} & $\sim$ 1100   &  $\frac{r_{\rm{s}}(z^*)}{D_{\rm{A}}(z^*)}$    & 1  \\
CMB - Primordial  &  \textit{Planck} 2018 TT, TE, EE + Lowl + lowE \citep{Planck}  & $\sim$ 1100   &  $A_{\rm{s}}$, $n_{\rm{s}}$     & 2  \\
 \hline
\end{tabular}
\tablefoot{We list the probe, the type of data selected, the redshift range of the probe, the choice of observable, and the size of the data vector. Expressions for the observables are provided in Sect.~\ref{Cosmological Observables}. A detailed description of the different data sets can be found in Sect.~\ref{Data Sets}.}
\end{table*}

To constrain the geometry and growth regimes of $\Lambda$CDM, we jointly analysed a range of recent measurements. A summary of the data sets employed in this work is given in Table~\ref{tab:data}.

\subsection{KiDS-1000 cosmic shear measurements} \label{K1K}
We employed cosmic shear measurements from the fourth data release of the European Southern Observatory's Kilo Degree Survey \citep[KiDS;][]{kuijken19} incorporating data from the fully overlapping VISTA Kilo-Degree Infrared Galaxy Survey \citep[VIKING; ][]{edge13}. The KiDS and VIKING surveys were designed to be complementary and combine optical and near-infrared imaging in nine photometric bands \citep{wright20}. We analysed the now-public weak lensing shear catalogue dubbed KiDS-1000 from \cite{giblin21}, which images 1006 ${\rm deg}^2$ on the sky \footnote{\url{http://kids.strw.leidenuniv.nl/DR4/lensing.php}}. This data set is divided into four photometric redshift bins of width $\Delta z = 0.2$ in the range $0.1 \leq z_{\rm B}\leq 0.9$ and a fifth bin with $0.9 \leq z_{\rm B} \leq 1.2$, based on their most probable Bayesian redshift $z_{\rm B}$ inferred with the code \textsc{BPZ} \citep{benitez00}. The redshift distributions of the five tomographic bins are then calibrated with deep spectroscopic samples that are reweighted using a self-organising map \citep{wright20, hildebrandt20}.

As summary statistic of the cosmic shear signal we adopted band power spectra estimated from the two-point correlation functions, which are analysed in \cite{K1K}. The corresponding KiDS-1000 cosmic shear likelihood is publicly available in the KiDS Cosmology Analysis Pipeline\footnote{\url{https://github.com/KiDS-WL/kcap}} (\textsc{KCAP}) together with a MontePython interface\footnote{\url{https://github.com/BStoelzner/KiDS-1000_MontePython_likelihood}} that wraps the \textsc{KCAP} functionality. This likelihood requires two additional astrophysical nuisance parameters: the amplitude of intrinsic galaxy alignments $A_{\rm IA}$ and the baryonic feedback parameter $A_{\rm bary}$;  see \citet{K1K_methodology} for further details.
Furthermore, five additional nuisance parameters $\delta_z$ allow for a shift of the mean of the redshift distribution in each tomographic bin within informative Gaussian prior set by the calibration procedure. Since these nuisance parameters do not have a cosmological interpretation we only kept one instance that was shared between the geometry and the growth instances of the cosmological code.

\subsection{Galaxy clustering} \label{Galaxy clustering }

The main data source we employed to draw constraints from the BAO and RSD observables were the Sloan Digital Sky Survey III \citep[SDSS III;][]{SDSS-III, BOSS} and the Six-Degree Field Galaxy Survey (6dFGS) \citep{6df_data, 6dF_BAO}.
Concerning the first of the two surveys, we made use of the 12th data release of the galaxy clustering data set of the Baryon Oscillation Spectroscopic Survey (BOSS DR12) which forms part of SDSS III. BOSS DR12 contains records of 1.2 million galaxies over an area of 9329 deg$^2$ and volume of 18.7 Gpc$^3$, divided into three partially overlapping redshift slices centred at effective redshifts 0.38, 0.51, and 0.61.

We fitted the geometrical relations $\alpha_\parallel$ and $\alpha_\perp$ (Eq. \ref{eq:alphas}) as reported by the BOSS DR12 data products:
 \begin{equation} \label{eq:alphas_B}
    \frac{[H(z)]_{\rm{fid}}}{\alpha_\parallel} = \frac{H(z) [r_{\rm{s}}(z_{\rm{d}})]_{\rm{fid}} }{ r_{\rm{s}}(z_{\rm{d}})}; ~~~
    \alpha_\perp [\chi(z)]_{\rm{fid}} = \frac{\chi / r_{\rm{s}}(z_{\rm{d}})}{[r_{\rm{s}}(z_{\rm{d}})]_{\rm{fid}}} ,
\end{equation}
from the reconstruction of the BAO feature at the three different redshift bins where $[r_{\rm{s}}(z_{\rm{d}})]_{\rm{fid}} = 147.78$ Mpc is the scale of the sound horizon at drag epoch as given by the fiducial cosmology used for the reconstruction. $[H(z)]_{\rm{fid}}$ and $[\chi(z)]_{\rm{fid}}$ are the corresponding Hubble parameter and comoving radial distance for the fiducial cosmology, respectively. When fitting these distance relationships, $r_{\rm{s}}(z_{\rm{d}})$ was treated as a free quantity to be determined by the fitting process while $[r_{\rm{s}}(z_{\rm{d}})]_{\rm{fid}}$ was fixed to the value provided in the data products. Moreover, we fitted the three redshift measurements of  $f\sigma_{\rm{8}}(z)$ from RSD obtained using the anisotropic clustering of the pre-reconstruction density field \citep{BOSS}. The BAO and RSD measurements of BOSS DR12 are two features extracted from the same set of observations. As such, they are not statistically independent. Thus, when analysing these two features, a combined analysis was performed mediated by the combined covariance matrix of the two sets of measurements from \citet{BOSS}. Moreover, we also accounted for the correlations present between the three BAO measurements with each other and between the three RSD measurements with each other resulting from the overlap in the used redshift bins.

This treatment of BOSS DR12 data given is different from \citet{KV+BOSS} and \citet{K1K+BOSS}, who performed a combined studies of KiDS  with BOSS DR12 in a full-shape analysis.
While performing a full-shape analysis in our study would increase the constraining power, it makes discerning between geometry and growth significantly more complex. Thus, we decided to trade constraining power for clarity in our proposed classification.

The 6dFGS is a combined redshift and peculiar velocity survey covering nearly the entire southern sky. The median redshift of the survey is z = 0.052. The 6dFGS BAO detection offers a constraint on the distance-redshift relation  $r_{\rm{s}}(z_{\rm{d}})/D_{\rm{v}}(z_{\rm{eff}})$ at $z_{\rm{eff}} = 0.106$. Altogether, we employed a total of 10 data points from clustering surveys. 

Both the SDSS and the 6dFGS overlap with KiDS in its northern and southern patches, respectively. While in principle this induces correlations with the weak lensing measurements, these are comfortably negligible, primarily because of the large survey areas outside the KiDS footprint used for clustering. Moreover, the measurements of $\alpha_\parallel$ and $\alpha_\perp$ tend to be extracted from larger physical scales than the weak lensing information. \citet{K1K_methodology} showed that cross-correlations between BOSS and KiDS are negligible even for a full-shape clustering analysis, and for the 6dFGS, which is at very low redshift, the correlation is even weaker.

\subsection{Lyman-$\alpha$ forest and quasars} \label{Lyman-a forest and quasars}

We employed high-redshift constraints on the BAO signature from SDSS-IV \citep{SDSS-IV} Data Release 14, observed as part of the eBOSS (\citealp{eBOSS_tech}). We did so by combining the auto- and cross-correlation analyses of three quasar samples from DR14Q \citep{ebossdata1, ebossdata2} which includes quasar clustering and  Ly$\alpha$ forest absorption in the Ly$\alpha$ and Ly$\beta$ regions. The selected sample of tracer quasars contains 266,590 quasars in the range $1.77 < z_{\rm{q}} < 3.5$. It includes 13,406 SDSS DR7 quasars \citet{SDSSDR7} and 18,418 broad absorption line (BAL) quasars \citep{Weymann}. The Ly$\alpha$ sample is derived from a super set consisting of 194,166 quasars in the redshift range $2.05 < z_{\rm{q}} < 3.5$, whereas the Ly$\beta$ sample is taken from a super set containing 76,650 quasars with $2.55 < z_{\rm{q}} < 3.5$.

The BAO signal is detected both parallel and perpendicular to the line of sight in the auto-correlation of the two quasar samples as well as in the cross-correlation between the two. This allows for the measurement of the distance relationships $\alpha_\parallel$ and $\alpha_\perp$ described in Eq.~(\ref{eq:alphas}), where the fiducial factors of normalisation are $[D_{\rm{M}}/r_{\rm{s}}(z_{\rm{d}})]_{\rm{fid}} = 39.26$ and $[D_{\rm{H}}/r_{\rm{s}}(z_{\rm{d}})]_{\rm{fid}} = 8.581$.
As opposed to the rest of the data sets, when fitting the Ly$\alpha$ data we did not assume a Gaussian likelihood for the distance relationships. Instead, we interpolated the publicly available MCMC chains of the combined analysis of Ly$\alpha$ auto-correlation, quasar auto-correlation and Ly$\alpha$-quasar cross-correlation\footnote{\url{https://github.com/brinckmann/montepython_public/blob/3.4/data/eBOSS_DR14_scans/eBOSS_DR14_Lya_combined_scan.dat}} from \citet{ebossdata1, ebossdata2} and evaluated the two-dimensional interpolation function at the sampled $\alpha_\parallel$ and $\alpha_\perp$ to obtain the corresponding $\Delta \chi^2$-value relative to the best-fit $\chi^2$-value; $\chi^2_{\rm{min}}$, obtained from \citet[table 5]{ebossdata2}. When reporting the goodness of fit obtained using this likelihood, we employed the following formula $\chi^2 = \Delta \chi^2 + \chi^2_{\rm{min}} = \Delta \chi^2 + 6499.31$; where $\Delta \chi^2$ is the value inferred with our likelihood code and $\chi^2_{\rm{min}} = 6499.31$ is the best goodness of fit found by \citet{ebossdata2}. When combining these data sets, since goodness of fit values are additive, we simply added $\chi^2_{\rm{min}}$ to the obtained $\Delta \chi^2$ of the combined data sets. 

\subsection{Cosmic microwave background anisotropies} \label{Cosmic microwave background anisotropies}

We made use of the constraints resulting from the analysis of measurements of the cosmic microwave background (CMB) temperature and polarisation anisotropy maps of the European Space Agency’s satellite \textit{Planck} \citep{Planck} denoted as TT, TE, EE + Lowl + lowE and referred to in the following as 'Planck 2018'. More specifically, we employed the posteriors on the cosmological parameters $A_{\rm{s}}$ and $n_{\rm{s}}$ as two data points. Additionally, we adopted \textit{Planck}'s posterior for the BAO angular scale $\theta^* = \frac{D_{\rm{A}}(z^*)}{r_{\rm{s}}*}$, where $z^* \sim 1100$ is the redshift of the end of the recombination epoch.   

In order to convert the previously mentioned posteriors into data points, we employed the \textsc{Python} Monte Carlo sample analysis package \textsc{GetDist} \citep{getDist} to first marginalise the public posterior chains of the \textit{Planck} 2018 TT, TE, EE + Lowl + lowE set over all parameters except $\theta^*$, $A_{\rm{s}}$ and $n_{\rm{s}}$. Then, we extracted the best-fit values of the parameters by finding the maximum of the joint posterior probability density distribution function.
Finally, we calculated the combined covariance matrix of the three parameters. In doing so, we approximated the marginalised posterior as a multivariate Gaussian, which is an accurate assumption (see e.g. Fig.~\ref{fig: CMB}). This allowed us to associate an error bar with each data point and to evaluate the level of correlation between the three pseudo-measurements.
This allowed us to build a Gaussian likelihood based on the three Planck constraints that we can use to inform our own parameter constraints. To distinguish this subset from the complete \textit{Planck} posterior we label it as Recomb
This methodology is analogous to that used by \citet{DES_geo_gro} when extracting constraints from a \textsc{MultiNest} chain of the TT + lowl \textit{Planck} lite 2015 likelihood and to the $\theta^*$ fitting process in \citet{Huterer} .

\section{Likelihood analysis} \label{Likelihood analysis}
 
Bayesian inference is employed to obtain constraints for the $\Lambda$CDM parameter sets governing the geometry and growth theory regimes: $\boldsymbol{p}^{\rm{geom}}$ and  $\boldsymbol{p}^{\rm{grow}}$, respectively. It relies on Bayes\rq\ theorem to relate the probability of a given set of parameter values conditioned on the observed data $\boldsymbol{D}$, known as the posterior probability $ P(\boldsymbol{p}_{\rm{grow}}, \boldsymbol{p}_{\rm{geom}}|\boldsymbol{D})$, to the probability of observing the data given a set of parameter values, known as the likelihood $\mathcal{L}(\boldsymbol{D}|\boldsymbol{p}_{\rm{grow}},\boldsymbol{p}_{\rm{geom}})$:  
\begin{equation}\label{eq:bayes}
 P(\boldsymbol{p}_{\rm{grow}}, \boldsymbol{p}_{\rm{geom}}|\boldsymbol{D}) = \frac{\mathcal{L}(\boldsymbol{D}|\boldsymbol{p}_{\rm{grow}},\boldsymbol{p}_{\rm{geom}}) \Pi(\boldsymbol{p}_{\rm{grow}}) \Pi(\boldsymbol{p}_{\rm{geom}})}{\mathcal{Z}(\boldsymbol{D})} \;.
\end{equation}
Here, we also defined the prior $\Pi$ for a set of model parameters and the evidence $\mathcal{Z}(\boldsymbol{D})$, which is independent of the parameters.

In line with previous analyses, we chose a Gaussian likelihood for all data sets under consideration,
\begin{multline} \label{eq:lkl}
 \mathcal{L}(\boldsymbol{D}|\boldsymbol{p}_{\rm{grow}},\boldsymbol{p}_{\rm{geom}}) \\
 = \frac{\exp \{ -\frac{1}{2}[\boldsymbol{D} - \boldsymbol{m}(\boldsymbol{p}_{\rm{grow}},\boldsymbol{p}_{\rm{geom}}) ]^TC^{-1}[\boldsymbol{D} -\boldsymbol{m}(\boldsymbol{p}_{\rm{grow}},\boldsymbol{p}_{\rm{geom}})]\}}{(2 \pi)^\frac{N}{2} \sqrt{\det(C)}} \;.
\end{multline}
In this general expression, $N$ is the dimension of the data vector, $C$ is covariance matrix that describes the statistical uncertainty and the correlations between the elements of $\boldsymbol{D}$, and  $\boldsymbol{m}(\boldsymbol{p}_{\rm{geom}},\boldsymbol{p}_{\rm{grow}})$ is the vector composed of the model predictions for the observations measured in $\boldsymbol{D}$, given the parameters. 
The weighted difference between the data and the model prediction is $\chi^2$-distributed for Gaussian data and will be used by us as a measure of the goodness of fit. 
 
 Bayesian inference takes into consideration the initial expectations for the values of the parameters prior to analysing the data via $\Pi(\boldsymbol{p})$. The priors chosen in this work for the parameters in both sets $\boldsymbol{p}^{\rm{geom}}$ and $\boldsymbol{p}^{\rm{grow}}$ are the same as in the original KiDS-1000 analysis \citep{K1K_methodology}, which were shown to bare no effect on the $S_{\rm{8}}$ posteriors. We display the prior distributions in Table~\ref{tab:prior}. We note that the KiDS cosmology priors are sufficiently conservative so as not to impact significantly on the posteriors of any of the probe combinations that we consider in this work. The table also provides a brief description of each parameter and specifies their role as either cosmological, nuisance or derived, i.e fully determined from the set of sampling parameters. We also show the treatment of each parameter as either split (i.e. present in both the geometry and the growth set of parameters) or shared (i.e. a single parameter was used to model both theory regimes).
 
The geometry and growth parametrisation entails a duplication of the cosmological parameter space and the associated prior volume, which affects model comparison and selection criteria \citep{Suspiciousness}. We note that alternative choices of priors could prove useful in this regard. For instance, instead of the sets $\boldsymbol{p}^{\rm{geom}}$ and $\boldsymbol{p}^{\rm{grow}}$ one could employ their mean and difference as the sampling parameters. While the mean would be assigned the same set of priors as the traditional analysis, the prior on the parameter differences could more explicitly account for our expectations in deviations from $\Lambda$CDM. We will consider these options in future work.
 
\begin{table*} 
\centering
\caption{Model parameters and their priors, adopted from \protect\citet{K1K}.}
\label{tab:prior}
\begin{tabular}{ lllll }
 \hline
 \hline
  Parameter  & Type & Prior & Duplicated & Description \\
    \hline
    $\omega_{\rm{cdm}}$ & Cosmological & $[0.051, 0.255]$ & $\checkmark$ & Reduced cold dark matter density parameter\\
    
    $S_{8}$       & Cosmological & $[0.1, 1.3]$ & $\checkmark$  & $S_{\rm{8}} = \sigma_{\rm{8}} \sqrt{\frac{\Omega_{\rm{m}}}{0.3}}$ \\
    
    $\omega_{\rm{b}}$   & Cosmological & $[0.019, 0.026]$ & $\checkmark$ & Reduced baryonic matter density parameter  \\
    
    $n_{\rm{s}} $        & Cosmological & $[0.84, 1.1]$ & $\checkmark$ & Spectral index of the primordial curvature power spectrum  \\
   
    $h$         & Cosmological & $[ 0.64, 0.82]$ & $\checkmark$ & Reduced Hubble parameter at present cosmic time  \\
    
    $\Omega_{\rm m}$  & Derived & - & $\checkmark$ & Total matter density parameter\\
    
    $\sigma_{8}$  & Derived & - & $\checkmark$ & Present-time amplitude of matter density fluctuations \\
    
    \hline
    
    $A_{\rm{IA}}$       &  Nuisance  & $[-6.0, 6.0]$ & $\times$  & Amplitude of intrinsic galaxy alignments\\
   
    $A_{\rm bary}$     & Nuisance & $[2, 3.13]$ & $\times$ & The baryonic feedback on the matter power spectrum  \\
    
    $\delta_z$         & Nuisance   &$\mathcal{N}\left(\boldsymbol{\mu},{\rm C}\right)$ & $\times$ & Shift of the mean of the KiDS redshift distributions\\
    \hline
    
    $\Omega_{\rm{k}}$  & Fixed &  0 & $\times$ & Curvature density parameter\\
    
    $\sum M_{\rm{ncdm}}$  & Fixed &  0.06 eV/$c^2$ & $\times$ & Total mass of massive neutrino species \\
    \hline
    
\end{tabular}
\tablefoot{The three sections correspond to cosmological, nuisance, and fixed parameters, respectively. The first column provides parameter names, with a brief description given in the fifth column. The second column lists whether parameters are sampled (distinguishing further between cosmological and nuisance parameters), fixed, or determined as a derived parameter from the posterior chain. The third column contains the fiducial parameter value (if fixed) or else the prior range, with the interval indicating the boundaries of a top-hat prior. The five shift parameters $\protect{\delta_z}$ are correlated through their covariance matrix C and their means $\protect \boldsymbol{\mu}$ are set to the mean values of the shifts of the tomographic redshift distributions from \protect\cite{hildebrandt20}. The fourth column shows whether a parameter is duplicated in the geometry--growth split or shared between the two regimes.}
\end{table*}

\begin{figure} 
\centering
    \includegraphics[width=.9\columnwidth]{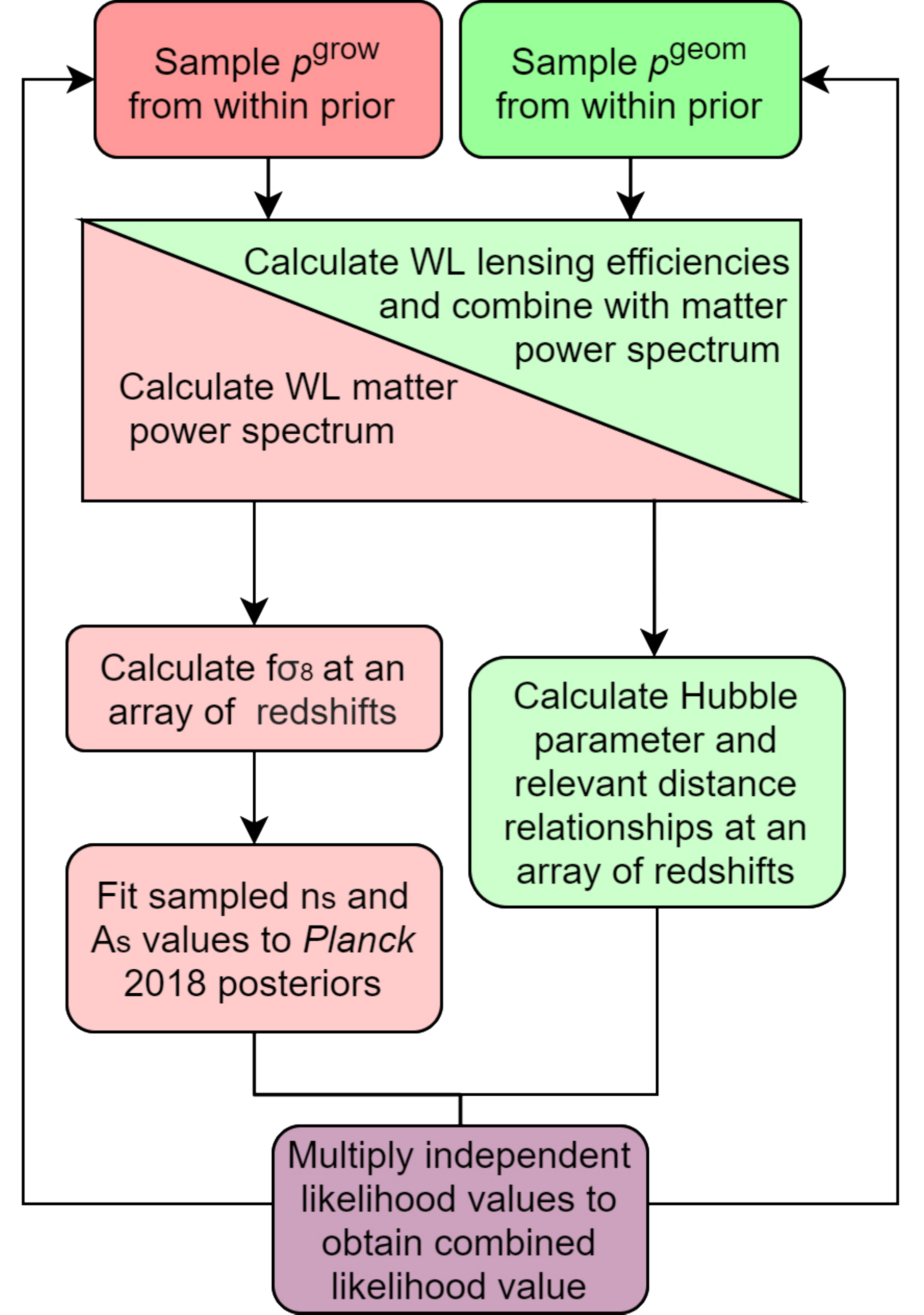} 
    \caption{Schematic of the complete \textsc{MontePython 2COSMOS} pipeline developed for the purposes of this work. Two instances of the cosmological code, cosmology 1 and cosmology 2, independently sample the cosmological parameters for the two theory regimes, geometry and growth, respectively. These parameters are then passed to the likelihood modules to calculate the theoretical prediction for each individual probe. Finally, the respective independent likelihood values for each probe are multiplied to obtain the likelihood value of the combined data set.}
    \label{fig:mark}
\end{figure}

The implementation of Bayesian inference relies on computer methods that make handling the typically high dimensionality of the parameter space feasible. This is especially true in this work where large parts of the parameter space have been doubled. In this work we made use of the cosmological parameter estimation code \textsc{MontePython 2COSMOS}\footnote{\url{https://github.com/fkoehlin/montepython_2cosmos_public}} \citep{Consistency}, a modification of \textsc{MontePython} \citep{Monte2, Monte1} that allows us to sample duplicated instances of parameters simultaneously. \textsc{MontePython 2COSMOS} achieves this by creating two separate instances of its underlying cosmological prediction code, \textsc{CLASS} \citep{blas11, Class}. Six pre-existing likelihoods from \textsc{MontePython} were adapted into \textsc{MontePython 2COSMOS} such that $\boldsymbol{p}^{\rm{grow}}$ and $\boldsymbol{p}^{\rm{geom}}$ can be assigned to the models of the observables discussed in Section~\ref{Cosmological Observables}; see Fig.~\ref{fig:mark} for a schematic overview. We employed the nested sampling algorithm \textsc{MultiNest} \citep{Multinest, Multinest2}, which allows us to reliably explore the high-dimensional posterior.

We assumed that the likelihoods for the different data sets employed are independent such that they can be combined by simple multiplication, $\mathcal{L}_{\rm tot}= \prod^N_i \mathcal{L}_i$, where $N$ is the number of probes used. The overlap region of the KiDS-1000 and BOSS footprints only accounts for 3$\%$ of the BOSS area, and hence it is safe to assume the two data sets are independent, as previous works have shown \citep{K1K_methodology}. Similarly, the BOSS and eBOSS data sets that we employed have been reported to be effectively independent \citep{Ata}. Also, the 6dfGS survey has been extensively used as an independent complement to SDSS data, as shown in \citet{6dF_BAO}.

The constraints of \textit{Planck} 2018 are known to be in tension with those of KiDS-1000 in $\Lambda$CDM and therefore should not be combined per se. This tension has also been shown to persist even when the $A_{\rm{s}}$ value is fixed \citep[see][]{K1K_beyond}. However, the tension does not manifest in the subset of parameter constraints that we adopted from \textit{Planck} (see Fig. \ref{fig: CMB} for reference). 
While the \textit{Planck} data points are independent of the low-redshift probes, joint posteriors employing this information will be necessarily correlated with full \textit{Planck} constraints that we display alongside.

\section{Results} 
\label{Results}

In this section we present constraints on the geometry and growth cosmological parameters combining the weak lensing, clustering, Lyman-$\alpha$, and Recomb data sets. When labelling the different results in tables and figures, results from the traditional, non-split analyses are simply labelled by the name of the combination of data sets employed, while labels of geometry and growth results also show the theory regime of the split model to which they belong. In Appendix~\ref{app:consistency},  we verify that these data sets are fully consistent with each other, and that our `traditional’ re-analysis of each data sets is consistent with the analyses presented in the literature.  We also analyse the goodness of fit for each data combination finding that in all studied cases the models offer a good fit of the data. Moreover, we compare the traditional and  geometry-growth split analysis, finding that while the split model is better at fitting the data, the improvement is not decisive at justifying the extra added degrees of freedom with respect the traditional analysis. We also calculate the Bayes factor and the deviance information criterion and find that none of the metrics display an statistically significant preference for either model for any combination of the data sets studied (see Appendix~\ref{app:Goodness of fit} for further details). 


\subsection{Geometry versus growth constraints}

\begin{figure*} 
\centering
    \includegraphics[width=1\linewidth]{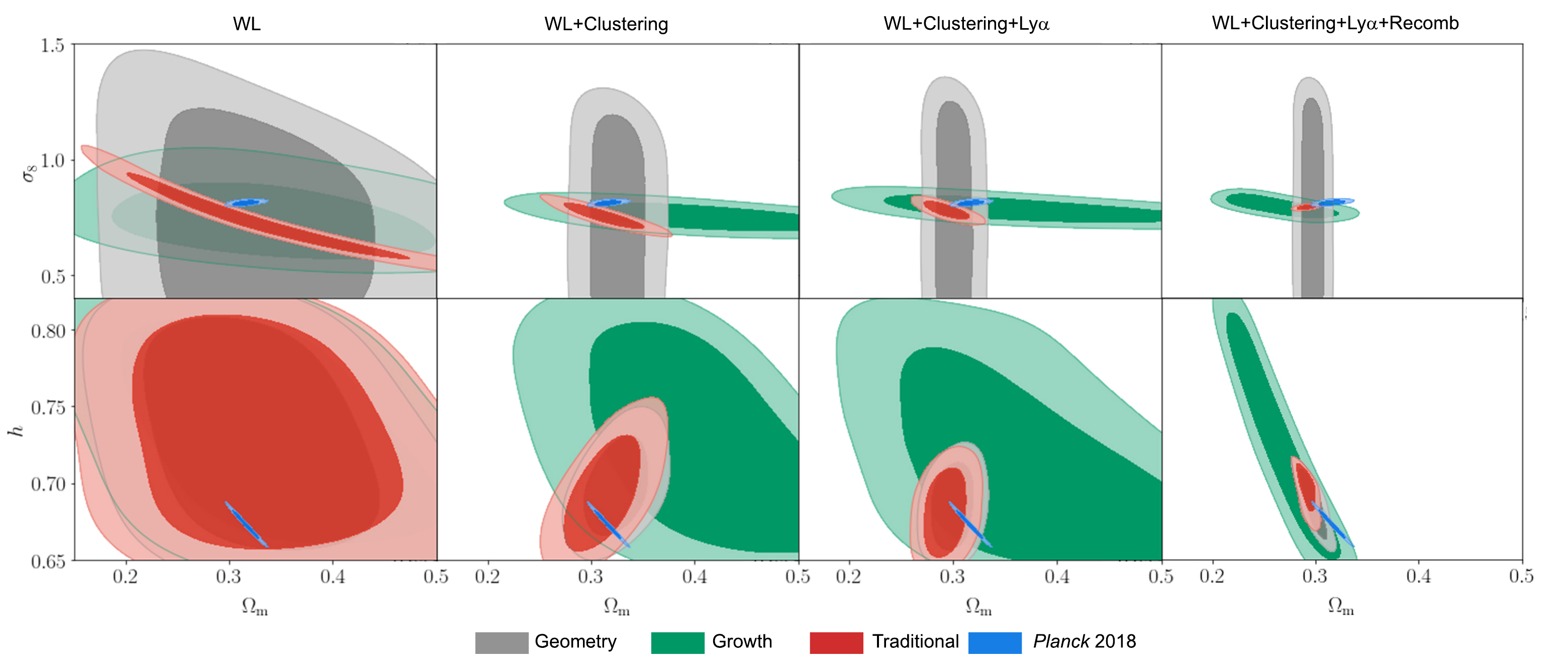} 
    \caption{
    Marginalised posterior distributions of $\sigma_{\rm{8}}$ and $\Omega_{\rm{m}}$ (top row), as well as $h$ and $\Omega_{\rm{m}}$ (bottom row) for different combinations of data sets (columns). Each panel shows a superposition of four contours. Namely, the growth and geometry contours from the split analysis of the two theory regimes (green and grey contours respectively), the contour resulting from the traditional analysis with one set of cosmological parameters (red), and the reference contours from the \textit{Planck} 2018 analysis \citep[blue;][]{Planck}.}
    \label{fig:2D_comp}
\end{figure*}

The marginalised posterior distributions for a subset of cosmological parameters, namely $\Omega_{\rm m}$, $\sigma_8$, and $h$, are illustrated in Fig. \ref{fig:2D_comp} for the split analysis of geometry (grey) and growth (green) as well as the traditional analysis with one single set of cosmological parameters (red). The leftmost row corresponds to an analysis of weak lensing data alone, while the rows to the right show the results obtained by subsequently adding the clustering, Lyman-$\alpha$, and Recomb data sets. Additionally, we display the constraints from the \textit{Planck} 2018 analysis \citep{Planck} for reference in each panel (blue). The corresponding marginalised one-dimensional  posteriors are shown in Fig.~\ref{fig:1D_comp} and the posteriors of the full set of cosmological parameters are presented in Appendix~\ref{app:Full Results}.

We find that the constraining power on the geometry and growth theory regimes differs depending on the sensitivity of the probes with respect to the various cosmological parameter in the two theory regimes. 
For the parameters $\Omega_{\rm m}$ or $\omega_{\rm{cdm}}$ the constraining power of geometry and growth is comparable. Potential discrepancies in these parameters would thus be the most meaningful as both regimes are informed by the data. 
We also observe cosmological parameters for which only one theory regime has a significant constraining power. This is the case for $S_{\rm{8}}$, $n_{\rm{s}}$ and $\sigma_{\rm{8}}$, that are only constrained by the growth regime. Consequently, the posteriors of their geometry counterparts are driven by the prior. Conversely, we find that $h$ is overwhelmingly dominated by the geometry regime while its growth posterior is only weakly informed by the role of background quantities in the calculation of growth quantities. However, this contribution to the $h^{\rm{gro}}$ parameter prevents it from simply returning its prior. Therefore, it is important to stress that while it is possible for the geometry regime to be  uninformed about the perturbatory cosmological parameters, the growth theory regime necessarily holds information, even if very limited, on the background parameters since the perturbatory aspects of the theory are built from this background cosmology.
Finally, we note that $\omega_{\rm{b}}$ is poorly constrained in both regimes with the data sets used. 

\begin{figure}
    \includegraphics[width=\linewidth]{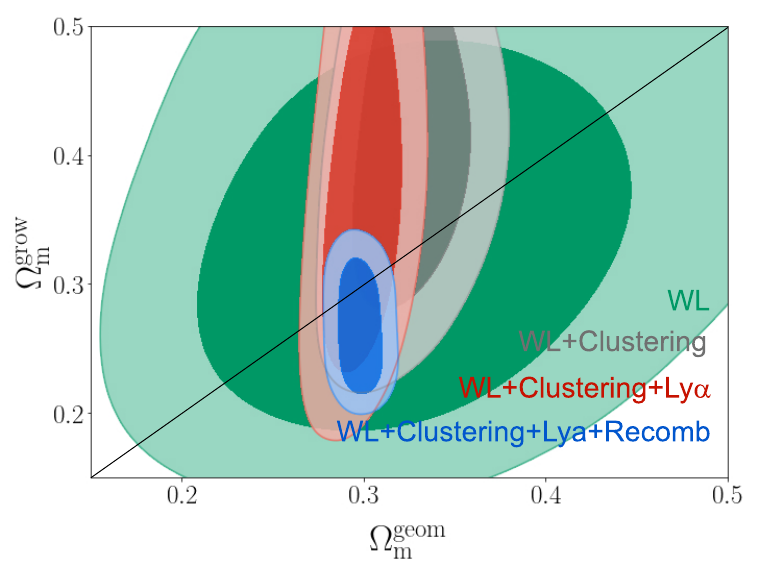}
    \caption{
    Marginalised posterior for $\Omega_{\rm{m}}$ when comparing their geometry (horizontal axis) and growth (vertical axis) counterparts. We show the evolution of the contours as more data sources are added into the analysis. Namely, we display weak lensing (green), weak lensing combined with clustering data (grey), weak lensing combined with clustering and Lyman-$\alpha$ forest data (red) and finally weak lensing combined with clustering, Lyman-$\alpha$ forest and Recomb data (blue).}
    \label{fig:comp_geo_gro}
\end{figure}

The geometry and growth regimes explore independent directions of the parameter space, which results in posterior distributions of $\Omega_{\rm m}$ and $\sigma_8$ that are orthogonal to each other. This was to be expected from the definition of each theory regime and the different information provided by geometry and growth observables. Figure~\ref{fig:2D_comp} shows that this orthogonality significantly strengthens as more data sets are combined. A lack of correlation between  $\boldsymbol{p}^{\rm{geom}}$ and $\boldsymbol{p}^{\rm{grow}}$ supports the meaningfulness of the geometry and growth categories as distinct theory regimes.
The constraints on $\Omega_{\rm m}$ from growth and on $\sigma_8$ from geometry are mainly dominated by the prior, except for the combination of weak lensing + clustering + Lyman-$\alpha$ + Recomb data, which puts stronger constraints from the growth regime on $\Omega_{\rm m}$. This causes the two regimes to span over two independent directions in parameter space. While $\Omega_{\rm{m}}^{\rm{grow}}$ is largely unconstrained, it is consistent with $\Omega_{\rm{m}}^{\rm{geom}}$ for all probe combinations, as detailed in Fig.~\ref{fig:comp_geo_gro}. This finding deviates somewhat from the similar study by \citet{DES_geo_gro} who found a $2\sigma$ preference for higher values of $\Omega_{\rm{m}}^{\rm{grow}}$ than $\Omega_{\rm{m}}^{\rm{geom}}$ in their combined probe analysis. The numerical values for the obtained $\Omega_{\rm{m}}$ constraints for each of the studied combinations of data sets can be found in Tab. \ref{tab:Omega_m}.

\begin{table} 
\centering
\caption{Marginal $\Omega_{\rm{m}}$ constraints.}
\label{tab:Omega_m}
\begin{tabular}{p{2.cm}ccc}
 \hline
 \hline
 \multicolumn{4}{c}{Marginal} \\
 \hline
 & & &   \\[-1em]
Data set & $\Omega_{\rm{m}}^{\rm trad}$ & $\Omega_{\rm{m}}^{\rm{geom}}$ & $\Omega_{\rm{m}}^{\rm{grow}}$ \\[0.5em] 

  \hline
& & &   \\[-1em]
WL  &  0.33$\pm$0.09 & 0.32$\pm$0.09 & 0.34$\pm$0.1 \\[0.5em] 
& & &   \\[-1em]
WL+Clustering   & 0.32$\pm$0.05 & 0.32$\pm$0.04 & 0.4$\pm$0.09  \\[0.5em] 
& & &   \\[-1em]
WL+Ly$\alpha$ & 0.25$\pm$0.08 & 0.24$\pm$0.07 & 0.30$\pm$0.1 \\[0.5em] 
& & &   \\[-1em]
WL+Recomb  & 0.28$\pm$0.03 & 0.26$\pm$0.05 & 0.29$\pm$0.06  \\[0.5em] 
& & &   \\[-1em]
WL+Clustering +Ly$\alpha$  & 0.30$\pm$0.05 & 0.30$\pm$0.04 & 0.38$\pm$0.1  \\[0.5em] 
& & &   \\[-1em]
WL+Clustering +Ly$\alpha$+Recomb & 0.29$\pm$0.02 & 0.29$\pm$0.02 & 0.27$\pm$0.04  \\[0.5em] 
   \hline

\end{tabular}
\tablefoot{The three columns display the obtained marginal constraints on $\Omega_{\rm{m}}$ in the form of mean $\pm$ standard deviation calculated using \texttt{GetDist}. In both sets of constraints the first column displays the traditional constraints, the middle column the constraints found by the geometry regime and last column those of the growth regime.}
\end{table}


Overall, we observe minimal or little correlation between the different pairs of cosmological parameters for any of the studied combinations of data sets, finding all pairs of cosmological parameter to have a correlation coefficient below $0.2$. The exception to this rule is the $\Omega_{\rm{m}}$ pair for which  we observe a positive correlation of 0.48 between  $\Omega^{\rm{geom}}_{\rm m}$- $\Omega^{\rm{grow}}_{\rm m}$ when solely analysing weak lensing data, that is sensitive to both theory regimes. The correlation vanishes once additional data sources are included as these put stronger constraints on $\Omega^{\rm geo}_{\rm m}$ while leaving $\Omega^{\rm gro}_{\rm m}$ mostly unchanged. 

The bottom row of Fig.~\ref{fig:2D_comp} shows the marginalised posterior distributions for $\Omega_{\rm{m}}$ and $h$. The weak lensing observable by itself is not sensitive to the Hubble parameter, which is why the posterior distributions for $h$ are prior-dominated for both geometry and growth, as well as for the traditional analysis. By adding additional data sets to the analysis, the geometry constraints shrink while the growth ones remain significantly wider, narrowing towards a $\Omega_{\rm m}h=\mbox{const}$ degeneracy constrained by the peak position of the matter power spectrum.

Comparing the posterior resulting from the traditional analysis to the ones obtained in the split analysis of the two theory regimes in the top row of Fig. \ref{fig:2D_comp}, we find that, as was expected, the traditional contours reside at the intersection between the geometry and growth contours. While the constraints on $\Omega_{\rm m}$ and $\sigma_8$ from geometry and growth by themselves are quite weak, the combination of the two regimes results in the 'banana-shaped' contour, which is most noticeable in the leftmost panel showing the constraints obtained from solely analysing weak lensing data.
The good agreement between the two theory regimes internally and the traditional constraints is also evidenced by the negligible change in the goodness of fit (see Table~\ref{tab:gof}).
However, tension with the full \textit{Planck} analysis remains throughout, which we will discuss further in Section~\ref{Cosmic tension}. 

\begin{figure} 
\centering
    \includegraphics[width=\linewidth]{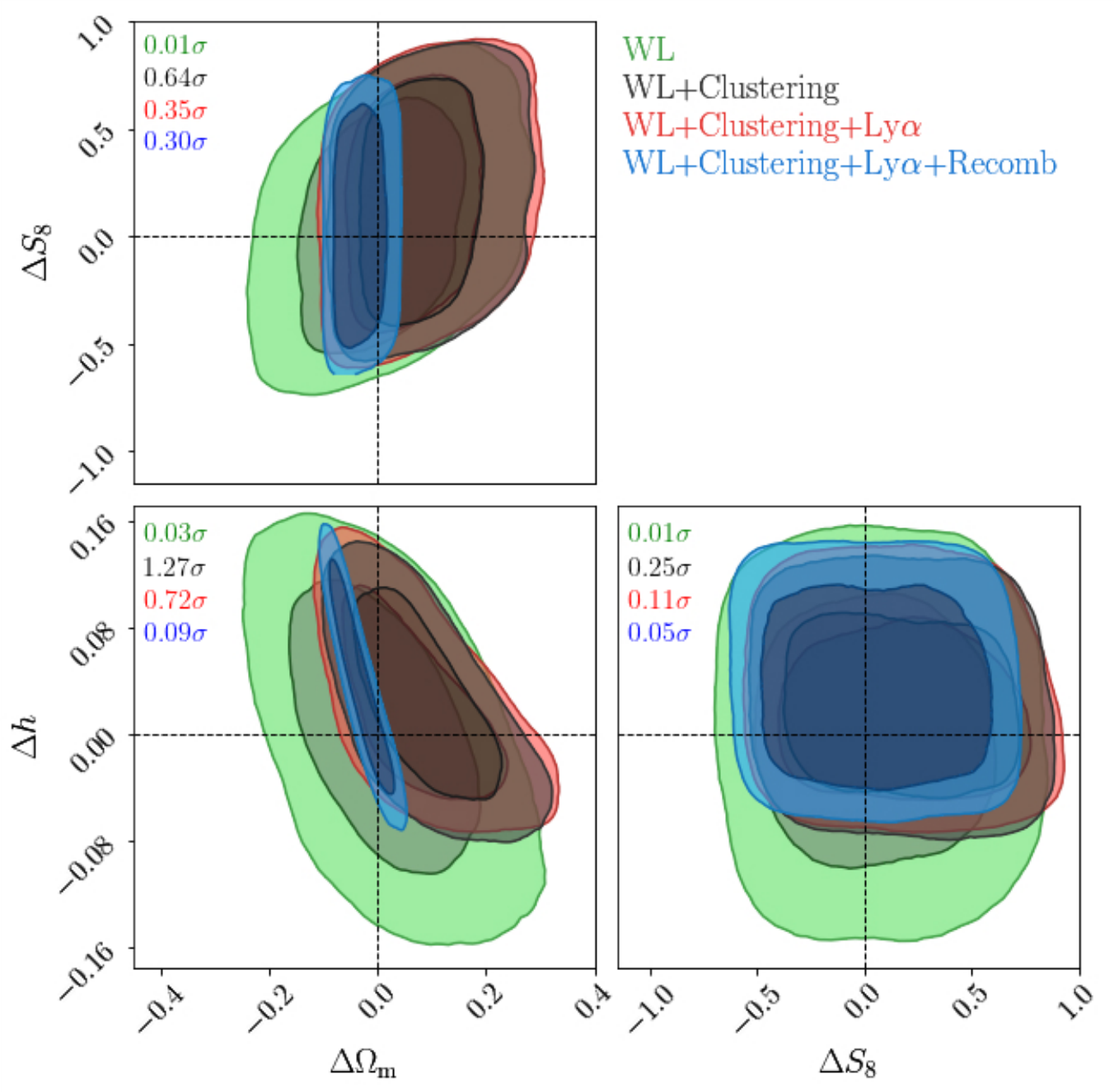} 
    \caption{
    Marginalised posterior distributions of the difference between geometry and growth parameter duplicates of $S_{\rm{8}}$, $\Omega_{\rm{m}}$, and $h$. Each panel shows the contours for the combinations of data sets WL (green), WL + Clustering (black), WL + Clustering + Ly$\alpha$ (red) and  WL + Clustering + Ly$\alpha$ + Recomb (blue). At the top-left corner of each panel we display the $\sigma$-level tension of each contour with respect to the null value $\boldsymbol{p}^{\rm{grow}}=\boldsymbol{p}^{\rm{geom}}$ for all combination of data sets.
    }
    \label{fig:diffs}
\end{figure}

We quantify the level of consistency between the two theory regimes by considering the posterior distribution of the difference between the parameter duplicates, which is shown in Fig.~\ref{fig:diffs} for three selected cosmological parameters: $S_{\rm{8}}$, $\Omega_{\rm{m}}$, and $h$. Visually, the marginalised posterior distributions show good agreement with the zero point. This indicates no tension between parameters in the two theory regimes. We quantify the tension between parameter duplicates following the methodology of \cite{Consistency}. We find good agreement between theory regimes for all combinations of parameters with a maximum offset of $1.27\sigma$ which is observed in the posterior distribution of $\Delta h$ and $\Delta \Omega_{\rm m}$ for weak lensing + clustering data.

Finally, we do not observe any new correlations between the different instances of the cosmological parameters and the nuisance parameters beyond those present in the traditional analysis. In particular, we do not see a correlation between the  growth instance of $\Omega_{\rm{m}}$ and the nuisance parameter $A_{\rm{IA}}$ as reported for weak lensing data by \citet{DES_geo_gro}.  However, it must be noted that this finding may be influenced by the fact that DESY1 and KiDS-1000 model the impact of intrinsic alignments differently.


\subsection{Tension with full \textit{Planck} data} 
\label{Cosmic tension}

The presence of tension between our  analyses and the full \textit{Planck} contours suggests that its cause may lie in the weak lensing observable, common to all cases of study, or in the $\Lambda$CDM model. The most recent cosmic shear analysis of KiDS-1000 \citep{K1K} found a 3$\sigma$ disagreement in their estimate of the cosmological parameter $S_{\rm{8}}$ with respect to the prediction of the \textit{Planck} 2018 analysis of the CMB.  This tension has been shown to persist both when combining the KiDS-1000 data with other probes \citep{K1K+BOSS} and when considering extensions of the $\Lambda$CDM model \citep{K1K_beyond}. 

The level of consistency between BOSS DR12 data on galaxy clustering \citep{BOSS} and \textit{Planck} 2018 depends on the chosen method of data compression \citep{Sanchez2017, Loureiro2019, Kobayashi2020}. \citet{KV+BOSS} showed that when employing the geometrical quantities $\alpha_\perp$ and $\alpha_\parallel$ the data are not in tension with the $\Lambda$CDM parameters of \textit{Planck} 2018. When cast into $\Lambda$CDM, the full shape analysis BOSS DR12 prefers a lower value of $\sigma_{\rm{8}}$ than \textit{Planck} 2018 at $2.1 \sigma$. However, when the tension is computed for the whole parameter space using the suspiciousness statistic \citep{Suspiciousness}, the two probes are in good agreement. 
The same considerations have to be made when assessing the consistency of the  eBOSS DR14 data set with other surveys. In the $\alpha_\perp$ and $\alpha_\parallel$ framework, the eBOSS DR14 has been shown to be consistent with the \textit{Planck} 2016 \citep{Planck16} best-fit flat $\Lambda$CDM model, with a mild deviation of $1.7\sigma$ \citep{ebossdata2}. However, a non-negligible degree of discrepancy; between 2 and 2.5 $\sigma$, has been reported when constraints are cast into the $\Lambda$CDM framework \citep{Aubourg15, Addison18}.

Finally, our implementation of the CMB data directly uses \textit{Planck} 2018 posteriors as data points ensuring a perfect agreement with the early-Universe probe. Interestingly, the analysis of Recomb data is in good agreement with that of WL (see Fig. \ref{fig: CMB}), suggesting that the cause of the tension between \textit{Planck} 2018 and WL must be driven by the features in the \textit{Planck} data over which we marginalise to create the Recomb data set.

Thus, of the four data sets considered in this work, only KiDS-1000 is in significant tension with \textit{Planck} 2018.  This is simply a consequence of the tension manifesting in the amplitude of structure growth for which the weak lensing data are most constraining.Nonetheless, this does not exclude the possibility of new tensions appearing upon the combination of data sets which independently are in good agreement with \textit{Planck} 2018; see the discussion for $\Omega_{\rm m}$ and $h$ below.

\begin{figure}
    \includegraphics[width=\linewidth]{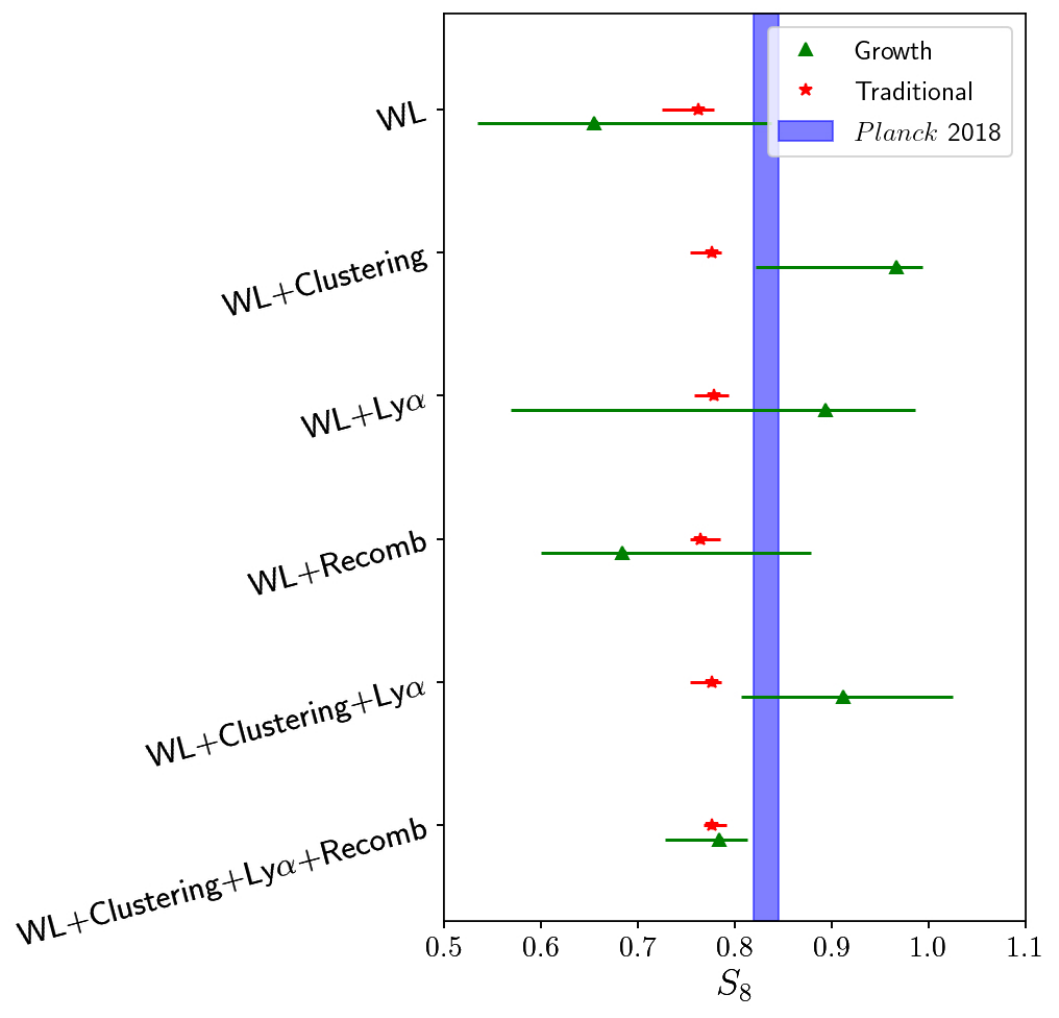}
    \caption{
     $S_{\rm{8}}$ best-fit parameter values with their associated 1$\sigma$  confidence regions obtained  from the different combinations of data sets explored in this work. In top to bottom order we display the data sets WL, WL + Clustering,  WL + Ly$\alpha$, WL + Recomb, WL + Clustering + Ly$\alpha$ and finally WL + Clustering + Ly$\alpha$ + Recomb  The numerical value of the quantities displayed can be found in Tab.~\ref{tab:S8}. }
    \label{fig:S8_comp}
\end{figure}

Since we use KiDS-1000 as our base observable in all combinations of data sets, we observe that the tension on $S_{\rm{8}}$  carries on to all our combinations of data sets. While the tension can already be appreciated in Figs.~\ref{fig:2D_comp} and \ref{fig:1D_comp}, we highlight this disagreement in Fig.~\ref{fig:S8_comp}, which shows the constraints on the parameter $S_{\rm{8}}$ for all the studied combinations of data sets, and for the traditional and growth parameters (the geometry constraints default to the prior and are not shown). We also show the maximum posterior (MAP) values for $S_{\rm{8}}$ and the associated 68\% credible interval (CI) calculated using its projected joint highest posterior density PJ-HPD (see \citealp{K1K_methodology} for reference) in Table~\ref{tab:S8}. 
The MAP values are directly inferred from the posterior distribution of sampling parameters. However, the limited number of samples in the posterior chains leads to some scatter in the MAP values, which is noticeable for the broader posterior in the weak lensing-only case. Therefore, we infer the MAP values for this specific data set using the Nelder-Mead optimisation method (\citealp{Nelder-Mead}; see \citealp{K1K_methodology} for details).

We explicitly quantify this tension under the assumption of Gaussian and independent marginal posteriors. In this case the tension $\tau$ between two data sets $i$ and $j$ for a parameter $p$ is given by
\begin{equation}
\label{eq:S8_tension}
    \tau_{\rm{ij}} = \frac{|\overline{p}_{\rm{i}} - \overline{p}_{\rm{j}}|}{\sqrt{{\rm Var}[p_{\rm{i}}]+{\rm Var}[p_{\rm{j}}]}}\;,
\end{equation}
where the barred quantities refer to the mean values of the distributions and Var to their variance. We provide the obtained values of the tension between the traditional analyses of the different combinations of data sets with \textit{Planck} 2018 for the parameter $S_{\rm{8}}$ in Table~\ref{tab:tension}. We see that the tension between the traditional and \textit{Planck} 2018 $S_{\rm{8}}$ posterior distributions stays within 2 to $3\sigma$ for all combinations of data sets.

Since each theory regime subset of parameters in our split model is a full $\Lambda$CDM model, it is possible to obtain the level of tension between geometry and growth with \textit{Planck} 2018. The \textit{growth} constraints are much weaker and hence not in tension although the combined constraint prefers a lower $S_8$ value than \textit{Planck}. We observe a slightly lower tension in $S_{\rm{8}}$ between our traditional KiDS-1000 only and \textit{Planck} 2018 than \citet{K1K} who employed a different summary statistic (COSEBIs) for their tension assessment. It is important to note that no particular choice of summary statistics is more powerful or more discrepant. Instead, the variance in the reported tension value for each summary statistic is due to slight differences in the degeneracy direction between the statistics which the parameter $S_{\rm{8}}$  does not perfectly capture. Moreover, our mean estimate is obtained directly from the posterior chains sampled with \textsc{MultiNest}, that is less accurate than the Nelder-Mead optimisation of \citet{K1K}.

\begin{figure}
    \includegraphics[width=\linewidth]{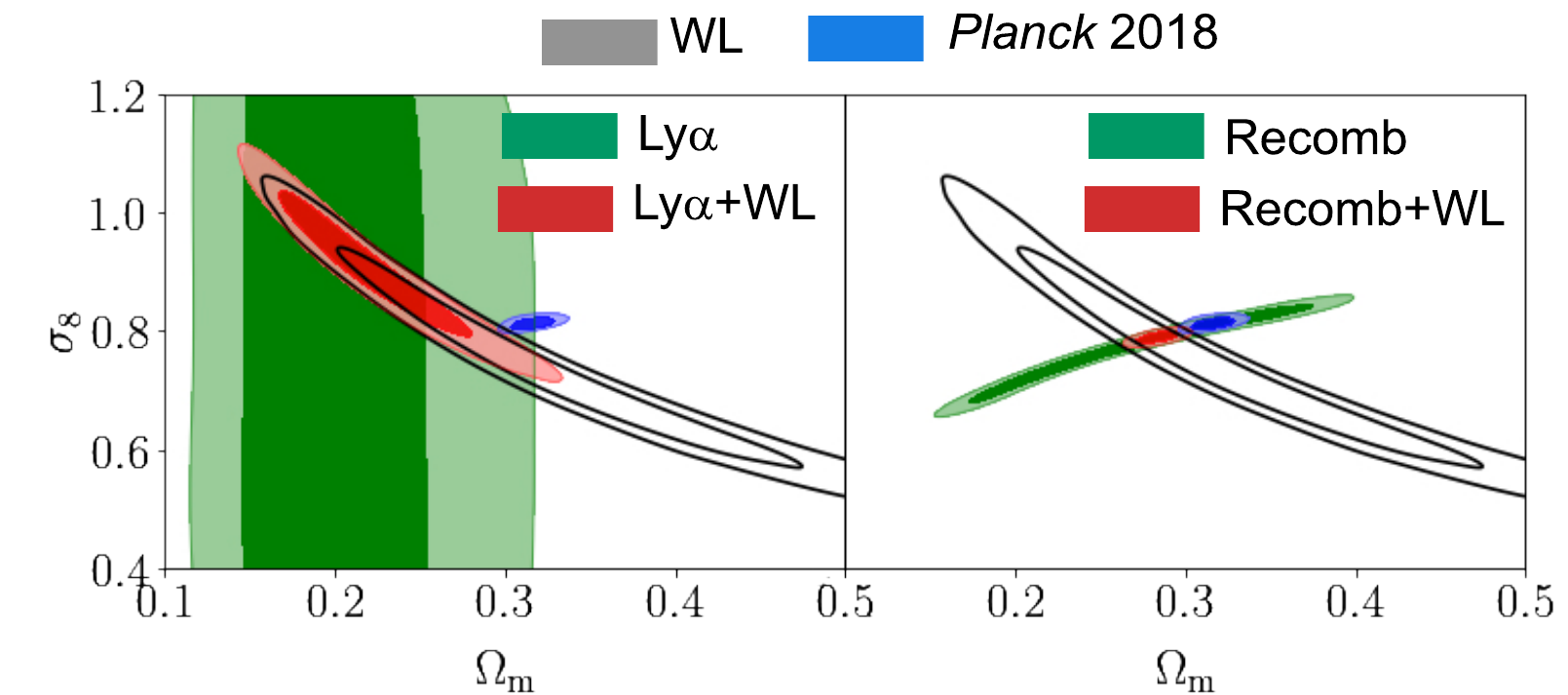}
    \caption{
    Two-dimensional constraints for the combination of parameters $\Omega_{\rm{m}}$-$\sigma_{\rm{8}}$ of the traditional analyses of Ly$\alpha$+WL data (left) and of Recomb+WL (right). In both panels the empty black contours are the WL alone constraints while the filled blue ones are the \textit{Planck} contours. In the left panel the green contours show the Ly$\alpha$ constraints while the red contours show their combination with WL. Similarly, in the right panel green contours show the Recomb constraints while the red contours show their combination with WL. The full posterior of these analyses can be seen in figures \ref{fig: Lya} and \ref{fig: CMB} respectively.}
    \label{fig: george}
\end{figure}

It is remarkable that the level of tension between traditional and \textit{Planck} contours increases by $0.3\sigma$ when our Recomb data (i.e. the acoustic peak angular scale and the primordial power spectrum parameters) is included in the analysis.  From Table~\ref{tab:S8}, it is possible to see Clustering and Lyman-$\alpha$ data push the $S_{\rm{8}}$ constraints towards higher values more affine to the full \textit{Planck} result, whereas Recomb; a subset of the CMB data,  has the opposite effect, which may seem counter-intuitive. However, we argue that this trend can be understood as a manifestation of the high versus low multipole discrepancy within \textit{Planck}  \citep[see Fig.~21 of][]{Planck}. Despite being a subset of the \textit{Planck} 2018 TTTEEE constraints for which there is no low versus high multipole discrepancy, Recomb is largely informed by the TT anisotropies within multipoles $\ell < 800$, and the low-$\ell$ TT \textit{Planck} posterior is in excellent agreement with our joint probe analysis.
Thus, it is possible to see how the combination of the Ly$\alpha$ data set preference for low values of $\Omega_{\rm{m}}$ in addition to the orthogonality between the degeneracy directions of the WL and Recomb contours increases the combined analysis tension. In Fig.~\ref{fig: george} an illustration of  the Ly$\alpha$ data set preference for low values of $\Omega_{\rm{m}}$ (left panel) and the orthogonality between the degeneracy directions of the WL and Recomb contours (right panel) is shown.

In contrast, the tentative signs of tension with the full \textit{Planck} constraints manifest not only in $S_8$ or $\sigma_8$, but extend to $\Omega_{\rm{m}}$ and $h$; as seen in Fig.~\ref{fig:2D_comp}. Discrepancy levels for $\Omega_{\rm{m}}$ are also given in Table~\ref{tab:tension} and reach $2.8 \sigma$ for the joint probe analysis including Recomb data and $2.4 \sigma$ in the WL + Clustering + Ly$\alpha$ + Recomb case. The Lyman-$\alpha$ forest data also has a preference for a somewhat lower $\Omega_{\rm{m}}$ than full \textit{Planck}, which translates into  a $2.3 \sigma$ discrepancy.
While the non-Gaussian shape of the marginal posterior for $h$ prevents us from applying Eq.~(\ref{eq:S8_tension}) more generally, in the WL + Clustering + Ly$\alpha$ + Recomb case the contours are sufficiently close to normal to apply the tension estimator, yielding a $2.0 \sigma$ tension in $h$. In agreement with the low-$\ell$ \textit{Planck} posterior, our joint probe posterior prefers lower values of $\Omega_{\rm{m}}$ and higher values of $h$ than full \textit{Planck}, with MAP values of $\Omega_{\rm{m}} \approx 0.289^{+0.007}_{-0.005}$ and $h \approx 0.705^{+0.007}_{-0.015}$.

It is important to bear in mind that our joint probe analysis including  the Recomb data is not statistically independent from the full \textit{Planck} 2018 constraints, violating the assumption made to obtain Eq.~(\ref{eq:S8_tension}). We leave a quantification of the level of correlation to future work. In principle, this correlation could lower or increase tension, dependent on the parameter degeneracies in the complement of the \textit{Planck} data that we did not employ in our analysis. However, since the \textit{Planck} high-$\ell$ constraints display the same degeneracy directions for the key parameters $\Omega_{\rm{m}}$, $\sigma_8$, and $h$ as the low-$\ell$ subset, it is reasonable to assume that this correlation is positive, and hence our tension estimates of Table~\ref{tab:S8} are lower bounds on the true level of discrepancy.

\begin{table*} 
\centering
\caption{Marginal $S_8$ constraints.}
\label{tab:S8}
\begin{tabular}{p{4.0cm}cccccc}
 \hline
 \hline
 & \multicolumn{3}{c}{MAP + PJ-HPD}&  \multicolumn{3}{c}{Marginal} \\
 \hline
 & & & & & \\[-1em]
Data set& $S_{\rm{8}}^{\rm trad}$ & $S_{\rm{8}}^{\rm{geom}}$ & $S_{\rm{8}}^{\rm{grow}}$ & $S_{\rm{8}}^{\rm trad}$ & $S_{\rm{8}}^{\rm{geom}}$ & $S_{\rm{8}}^{\rm{grow}}$ \\[0.5em] 

  \hline
& & & & & \\[-1em]
WL  &   $0.762_{-0.037}^{+0.016}$  & $0.704_{-0.300}^{+0.402}$ & $0.654_{-0.120}^{+0.183}$ & 0.751$\pm$0.027 & 0.719$\pm$0.310 & 0.779$\pm$0.139 \\[0.5em] 
& & & & & \\[-1em]
WL + Clustering  & $0.776_{-0.022}^{+0.010}$ & $0.676_{-0.184}^{+0.610}$ & $0.966_{-0.145}^{+0.027}$ & 0.768$\pm$0.016 & 0.700$\pm$0.326 & 0.871$\pm$0.077  \\[0.5em] 
& & & & & \\[-1em]
WL + Ly$\alpha$  & $0.778_{-0.021}^{+0.015}$ & $0.373_{-0.116}^{+0.639}$  & $0.893_{-0.324}^{+0.324}$ & 0.772$\pm$0.017 & 0.700$\pm$0.326 & 0.902$\pm$0.160 \\[0.5em] 
 & & & & & \\[-1em]
WL + Recomb & $0.764_{-0.011}^{+0.021}$ & $0.762_{-0.282}^{+0.505}$ & $0.684_{-0.084}^{+0.191}$ & 0.768$\pm$0.016 & 0.701$\pm$0.328 & 0.828$\pm$0.131  \\[0.5em] 
& & & & & \\[-1em]
WL + Clustering + Ly$\alpha$  & $0.777_{-0.023}^{+0.009}$ & $0.375_{-0.140}^{+0.627}$ & $0.911_{-0.105}^{+0.114}$ & 0.769$\pm$0.016 & 0.695$\pm$0.328 & 0.867$\pm$0.091  \\[0.5em] 
  & & & & & \\[-1em]
WL + Clustering + Ly$\alpha$ + Recomb & $0.776_{-0.008}^{+0.016}$ & $0.468_{-0.261}^{+0.533}$ & $0.783_{-0.056}^{+0.029}$ & 0.781$\pm$0.012 & 0.702$\pm$0.336 & 0.748$\pm$0.035  \\[0.5em] 
   \hline

\end{tabular}
\tablefoot{The first three columns display the maximum posterior (MAP) values for $S_{\rm{8}}$ and its 68\% credible interval (CI) calculated using its projected joint highest posterior density (PJ-HPD). The last three columns display the obtained marginal constraints on $S_{\rm{8}}$ in the form of mean $\pm$ standard deviation calculated using \texttt{GetDist}. In both sets of constraints the first column displays the traditional constraints, the middle column the constraints found by the geometry regime and last column those of the growth regime.}
\end{table*}

\begin{table} 
\centering
\caption{Level of tension between traditional analysis and \textit{Planck}} for the parameters $S_{\rm{8}}$ and  $\Omega_{\rm{m}}$.
\label{tab:tension}
\begin{tabular}{p{5cm}ll}
 \hline
 \hline
 Tension Trad. versus \textit{Planck} &  $\tau_{S_{\rm{8}}}$ & $\tau_{\Omega_{\rm{m}}}$ \\
  \hline
WL  &  2.62 &  0.19\\ 
WL + Clustering  & 2.65 & 0.11  \\ 
WL + Ly$\alpha$  & 2.64 & 2.28\\ 
WL + Recomb &  2.65$^*$ & 2.82$^*$ \\
WL + Clustering + Ly$\alpha$  & 2.65 & 1.43 \\
WL + Clustering + Ly$\alpha$ + Recomb & 2.91$^*$ &  2.40$^*$ \\
 
  \hline
\end{tabular}
\tablefoot{Tension was calculated using the method of difference of Gaussians between the traditional constraints and those of \textit{Planck} 2018 \citep{Planck}. The entries marked by a star employ a subset of \textit{Planck} CMB data in the probe combination and are therefore not statistically independent.}
\end{table}

\section{Conclusions} \label{Conclusions}

In this work we developed a multi-probe self-consistency test of the spatially flat $\Lambda$CDM model with the aim of exploring potential causes of the cosmic tension within the current cosmological theory. In order to do so, we divided our model into two theory regimes, geometry and growth, distinguishing between the strictly uniform background and the formation of matter density fluctuations on top of this background, respectively.  Making use of this distinction, we classified a series of cosmological observables, or parts thereof, as geometry or growth depending on the theory regime needed to model them within the $\Lambda$CDM model. We duplicated the $\Lambda$CDM parameter space into two independent copies, $\boldsymbol{p}^{\rm{grow}}$ and $\boldsymbol{p}^{\rm{geom}}$, and let each govern its respective set of observables.

As cosmological observables, we employed weak lensing (WL) cosmic shear measurements from the latest data release of the Kilo Degree Survey (KiDS-1000), and measurements of the galaxy and Lyman-$\alpha$ (Ly$\alpha$) BAO feature distance relationship from the 12th data release of the  Baryon Oscillation Spectroscopic Survey (BOSS DR12), as well as from the 14th data release (DR14) of the eBOSS and the 6 Degree Field Galaxy Survey (6dFGS). We also used growth rate measurements from BOSS DR12. Moreover, we made use of the \textit{Planck} 2018  posterior distributions for the cosmological parameters $A_{\rm{s}}$ and $n_{\rm{s}}$ and the angular scale of the sound horizon $\theta^*$ as pseudo-data points. We grouped these observables into four data sets: WL composed of the KiDS-1000 data, clustering which combined BOSS DR12 and the 6dfGS galaxy clustering measurements, Ly$\alpha$ composed of eBOSS DR14 data, and Recomb, containing the  subset of the \textit{Planck} 2018 posterior.
Constraints for $\boldsymbol{p}^{\rm{grow}}$ and $\boldsymbol{p}^{\rm{geom}}$ were obtained using a modified version of the Bayesian parameter estimation code \textsc{MontePython 2COSMOS}. In order to explore the extended parameter space we made use of Monte Carlo Markov chains while employing the nested sampler \textsc{MultiNest}. The code developed to perform this analysis is made publicly available \footnote{\url{https://github.com/BStoelzner/KiDS_geometry_vs_growth}}. 

We generally found very good agreement between all probes considered, including a subset of CMB data we used, with little variation in the goodness of fit. The geometry and growth parameters are  consistent throughout, and the additional degrees of freedom in the model due to the split are not preferred by the data. The constraints on $\boldsymbol{p}^{\rm{grow}}$ and $\boldsymbol{p}^{\rm{geom}}$ converge towards the those found by the traditional $\Lambda$CDM analysis that does not duplicate parameters for all the studied combinations of data sets. We also observed that $\boldsymbol{p}^{\rm{grow}}$ and $\boldsymbol{p}^{\rm{geom}}$ explore different directions of the parameter space, their contours tending to be orthogonal to each other. Thus, we conclude that our constraints on $\boldsymbol{p}^{\rm{grow}}$ and $\boldsymbol{p}^{\rm{geom}}$ support the geometry versus growth distinction as a meaningful classification of the $\Lambda$CDM model.
    
Regarding the $S_{\rm{8}}$ parameter tension between low- and high-redshift probes, our analysis produced tension levels between 2 and $3\sigma$ for the traditional constraints of different probe combinations and \textit{Planck} 2018 \citep{Planck}. The joint probe growth constraint on $S_8$ also prefers lower values, but is not in tension due to substantially larger statistical errors. The traditional constraints of the parameters $\Omega_{\rm{m}}$ and $h$ also reach discrepancies in the 2 and $3\sigma$ range, with larger values preferred for $h$ and smaller values preferred for $\Omega_{\rm{m}}$, in particular when our subset of CMB data is included. As this subset is primarily informed by large angular scales in the Recomb, this result supports earlier indications that the cosmic tensions in $S_8$, and possibly also $H_0$, are driven by multipoles $\ell > 800$ in \textit{Planck} \citep{p15_multipoles, Planck}.

In comparison to the recent similar analysis by \citet{DES_geo_gro}, our work employs an alternative distinction between geometry and growth solely based on the need to consider matter anisotropies in the modelling of the cosmological observable, irrespective of whether the matter anisotropies reside in the present or early Universe. This leads to a different classification  of cosmological observables between the two works. Nonetheless, both \citet{DES_geo_gro} and this work report an excellent degree of agreement between the two considered regimes. However, we do not observe the slight preference for higher values of $\Omega_{\rm{m}}^{\rm{grow}}$ as seen by \citet{DES_geo_gro}, nor do we find any significant correlations between $\Omega_{\rm{m}}^{\rm{grow}}$  and the shared nuisance parameter for the intrinsic alignment amplitude, $A_{\rm{IA}}$.

To conclude we outline possible avenues for improvements that future iterations of the methodology presented in this work could consider.
On the theoretical side, it would be interesting to study the feasibility of geometry versus growth splits beyond $\Lambda$CDM. It is clear that the distinction proposed in this work can be extrapolated straightforwardly to modified models that only introduce changes to the background cosmology, such as non-flat $\Lambda$CDM models or dark energy models with an effective equation of state, which have proven to be successful at encapsulating a wide variety of models \citep{dina_w0wa, Carlos_w0wa}. Another type of modification that is compatible with this framework is a time-varying Newton's constant for gravity because it does not change the shape of the Jeans equation \citep{baker_ferreira}.  Other, more involved modifications could include scale-dependent modifications to the growth of structure. More general parametrisations than a simple duplication of $\Lambda$CDM parameters would have to be considered, and the link between the background and the growth of structure, via the Jeans equation and its non-linear corrections, carefully investigated. 

The geometry and growth split is complementary to other empirical extensions to the standard $\Lambda$CDM model, such as $\mu$-$\Sigma$ models \citep{Pogosian16}. Theses models test for the consistency of the predictions of general relativity for relativistic and non-relativistic matter, whereas our split analysis explores the consistency between the predictions for the expansion history and structure growth.
As opposed to previous works that subdivided the $\Lambda$CDM parameter space between geometry and growth, in this work we fully duplicated the parameter space; assigning one complete set of $\Lambda$CDM parameters to each theory regime. This was done with the aim of ensuring the independence of the constraints of each theory regime and of adequately capturing the uncertainty induced by the limited information available to each regime. Reflecting on this choice of analysis, geometry's complete lack of constraining power on parameters that exclusively regards the anisotropies in the matter density field  relegates the geometry instances of such parameters as simple measures of uncertainty. Therefore, future analyses could avoid sampling over said parameters in the geometry regime and use the assigned prior distributions as the measures of uncertainty when deriving quantities of interest.

Future applications should strive to incorporate new measurements on similar or different physical observables that could be added as new sources of constraining power. Specially, efforts should be directed towards performing a full-shape analysis of the LSS matter power spectrum as shown in \citet{KV+BOSS} and \citet{K1K+BOSS} instead of the BAO and RSD feature analysis used in this work. Similarly, the current treatment of CMB data should be replaced by a more rigorous geometry and growth analysis of the full CMB data set, complemented by an analogous split of the CMB lensing likelihood \citep{CMBlensing} observable equivalent to the one undertaken for KiDS-1000.

In addition to this, we recommend updating the galaxy clustering data to the latest eBOSS DR16 release \citep{BOSSDR16}, which also includes $\sigma_{\rm{8}}$ at high redshifts  from quasar density measurements as well as high-redshift galaxy measurements \citep{DR16_galaxies}. Particularly adding to the currently weaker growth constraints, future works could consider studying the addition of Sunyaev-Zeldovich measurements of cluster counts \citep{P15_Zel}, as well as the Integrated Sachs-Wolfe effect \citep{ISW, stolzner18}. The geometry versus growth approach holds promise to yield novel insights with the powerful data from forthcoming surveys, including DESI \footnote{\url{https://www.desi.lbl.gov/}} \citep{DESI}, Euclid \footnote{\url{https://www.euclid-ec.org/}} \citep{Euclid}, and the LSST at the Rubin Observatory  \footnote{\url{https://www.lsst.org/}} \citep{LSST}.

\section*{Acknowledgements}

We thank Carlos Garcia-Garcia, David Alonso, and Shahab Joudaki for useful discussions. 
JRZ is supported by an STFC doctoral studentship and a Royal Astronomical Society summer student bursary.
BS and BJ acknowledge support by the UCL Cosmoparticle Initiative.
MA, BG, CH, and TT acknowledge support from the European Research Council under grant number 647112. MB is supported by the Polish National Science Center through grants no. 2020/38/E/ST9/00395, 2018/30/E/ST9/00698 and 2018/31/G/ST9/03388,
and by the Polish Ministry of Science and Higher Education through
grant DIR/WK/2018/12. AD, HH, JLvdB, and AHW are supported by an European Research Council Consolidator Grant (No. 770935). BG also acknowledges support from the Royal Society through an Enhancement Award (RGF/EA/181006). CH also acknowledges support from the Alexander von Humboldt Foundation in the framework of the Max Planck-Humboldt Research Award endowed by the Federal Ministry of Education and Research. HH is also supported by a Heisenberg grant of the Deutsche Forschungsgemeinschaft (Hi 1495/5-1). KK acknowledges support from the Royal Society and Imperial College. TT also acknowledges funding from the European Union’s Horizon 2020 research and innovation programme under the Marie Sk\l{}odowska-Curie grant agreement No.~797794.
The analysis made use of the software tools
\textsc{SciPy} \citep{scipy}, \textsc{NumPy} \citep{np}, \textsc{Matplotlib} \citep{np},
\textsc{CLASS} \citep{Class}, GetDist \citep{getDist}, \textsc{MultiNest} \citep{Multinest, Multinest2},
\textsc{MontePython} \citep{Monte1, Monte2}
and \textsc{MontePython 2COSMOS} \citep{Consistency}.

Based  on  observations  made with ESO Telescopes at the La Silla Paranal Observatory under programme IDs 177.A-3016, 177.A-3017, 177.A-3018, 179.A-2004, 298.A-5015, and on data products produced by the KiDS consortium.
Based on observations obtained with \textit{Planck} (\url{http://www.esa.int/Planck}), an ESA science mission with instruments and contributions directly funded by ESA Member States, NASA, and Canada.
We acknowledge the efforts of the staff of the Anglo-Australian Observatory, who have undertaken the observations and developed the 6dF instrument.

Funding for the Sloan Digital Sky Survey IV has been provided by the  Alfred P. Sloan Foundation, the U.S. Department of Energy Office of Science, and the Participating Institutions. SDSS-IV acknowledges support and  resources from the Center for High Performance Computing  at the University of Utah. The SDSS website can be found at \url{www.sdss.org}.
SDSS-IV is managed by the  Astrophysical Research Consortium for the Participating Institutions of the SDSS Collaboration including  the Brazilian Participation Group,  the Carnegie Institution for Science,  Carnegie Mellon University, Center for Astrophysics | Harvard \& Smithsonian, the Chilean Participation, the French Participation Group,  Instituto de Astrof\'isica de Canarias, The Johns Hopkins  University, Kavli Institute for the Physics and Mathematics of the Universe (IPMU) / University of Tokyo, the Korean Participation Group, Lawrence Berkeley National Laboratory, Leibniz Institut f\"ur Astrophysik Potsdam (AIP),  Max-Planck-Institut f\"ur Astronomie (MPIA Heidelberg), Max-Planck-Institut f\"ur  Astrophysik (MPA Garching), Max-Planck-Institut f\"ur Extraterrestrische Physik (MPE), National Astronomical Observatories of China, New Mexico State University, New York University, University of Notre Dame, Observat\'ario Nacional / MCTI, The Ohio State University, Pennsylvania State University, Shanghai Astronomical Observatory, United Kingdom Participation Group, Universidad Nacional Aut\'onoma de M\'exico, University of Arizona, University of Colorado Boulder, University of Oxford, University of Portsmouth, University of Utah, University of Virginia, University of Washington, University of Wisconsin, Vanderbilt University, and Yale University.

\textit{Author contributions}: All authors contributed to the development and writing of this paper. The authorship list is given in two groups: the lead authors (JRZ, BS, BJ), followed by an alphabetical group of co-authors who have either made a significant contribution to the data products or to the scientific analysis.




\bibliographystyle{aa}
\bibliography{aanda}



\appendix
\section{Consistency of data sets}
\label{app:consistency}

\begin{figure*}[p] 
        \centering
        \includegraphics[width=\linewidth]{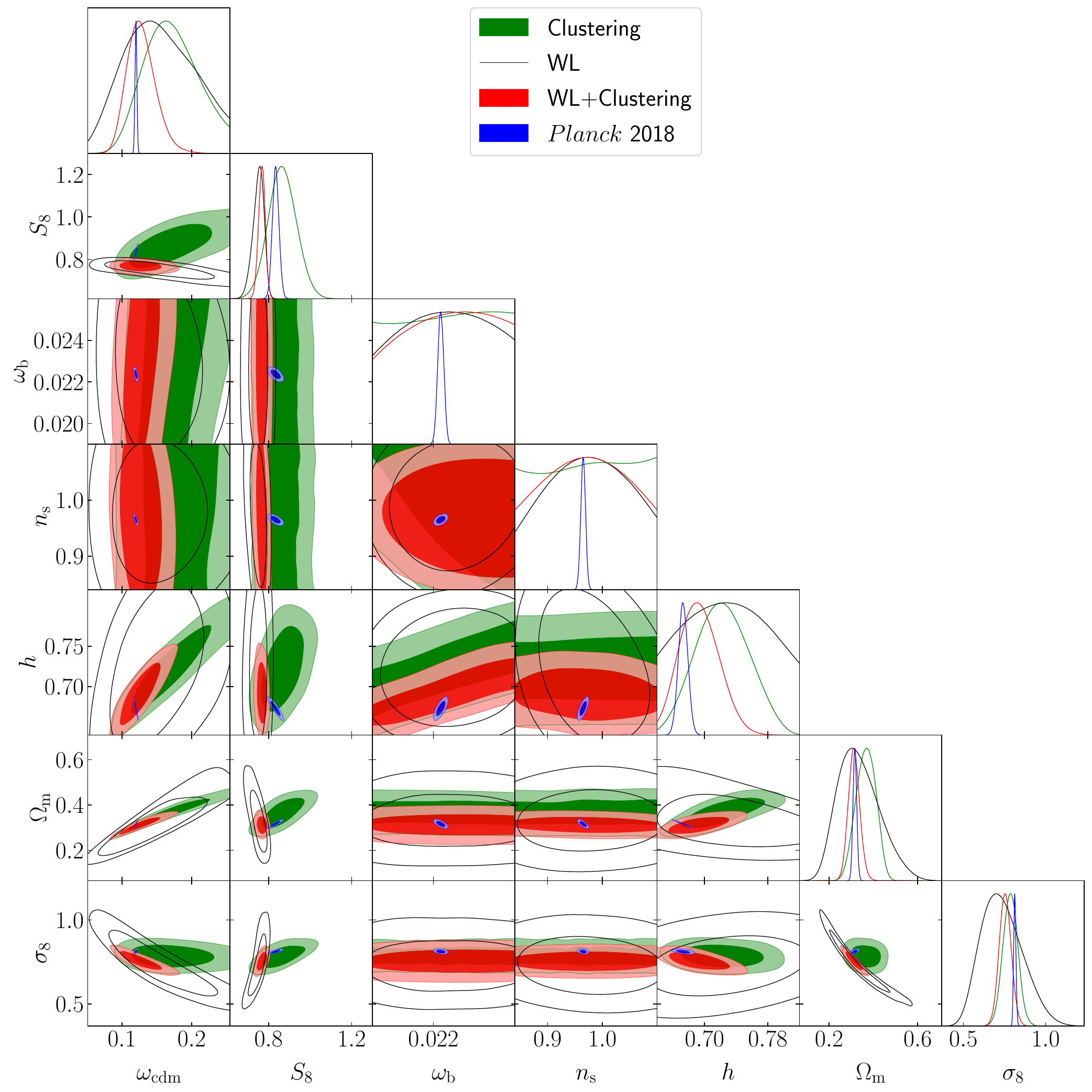} 
        \caption{
        Marginalised posteriors distributions from the analysis of the WL and Clustering data sets. Each panel shows the posterior resulting from a separate analysis of clustering (green) and weak lensing (black) data sets as well as the combination of both data sets (red). Additionally, we show the \textit{Planck} 2018 contours \citep{Planck} for reference (blue).}
        \label{fig: BOSS}
        \end{figure*}

\begin{figure*}[p] 
        \centering
        \includegraphics[width=\linewidth]{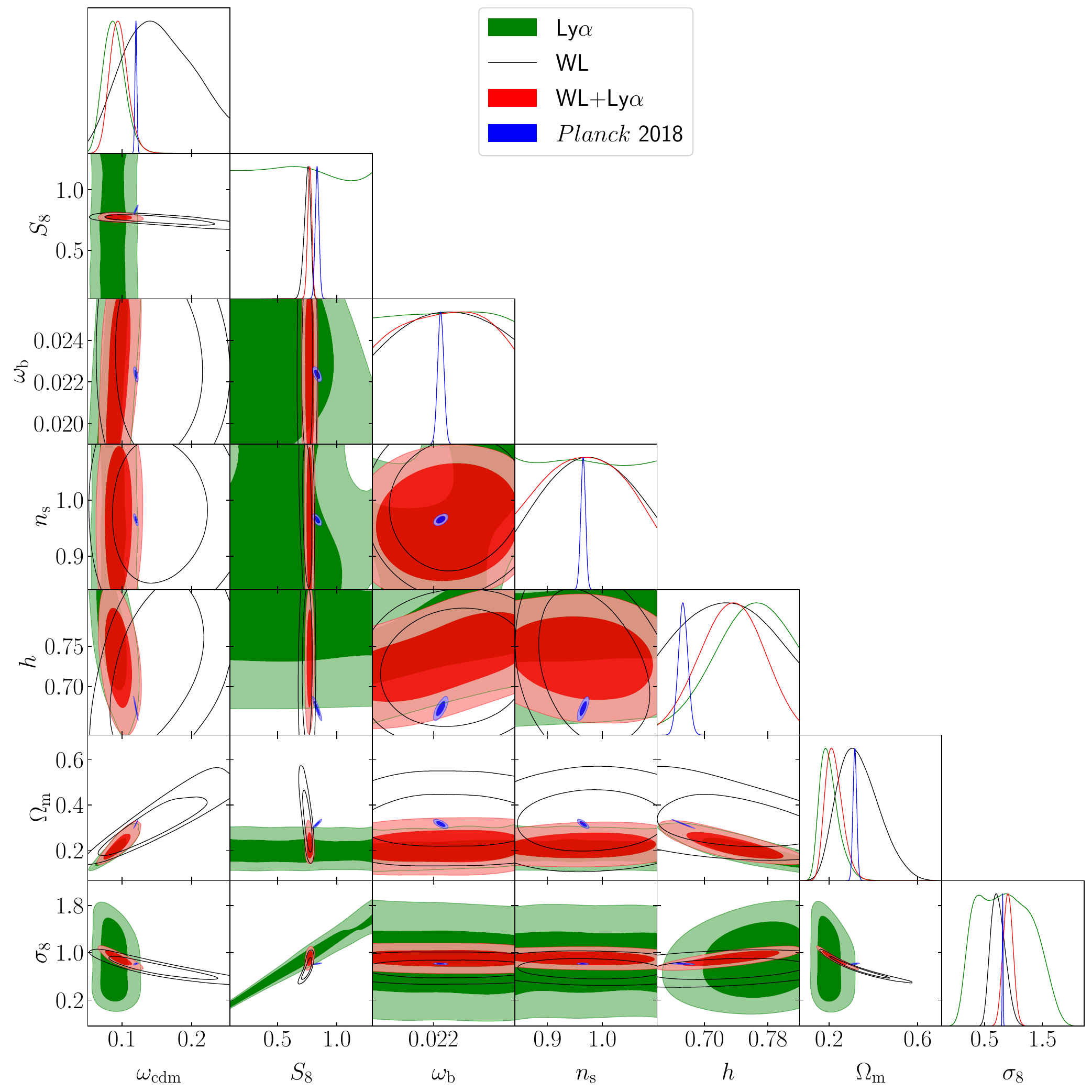} 
        \caption{
        Marginalised posteriors distributions from the analysis of the WL and Ly-$\alpha$ data sets. Each panel shows the posterior resulting from a separate analysis of Lyman-$\alpha$ (green) and weak lensing (black) data sets as well as the combination of both data sets (red). Additionally, we show the \textit{Planck} 2018 contours \citep{Planck} for reference (blue).}
        \label{fig: Lya}
        \end{figure*}

\begin{figure*}[p] 
        \centering
        \includegraphics[width=\linewidth]{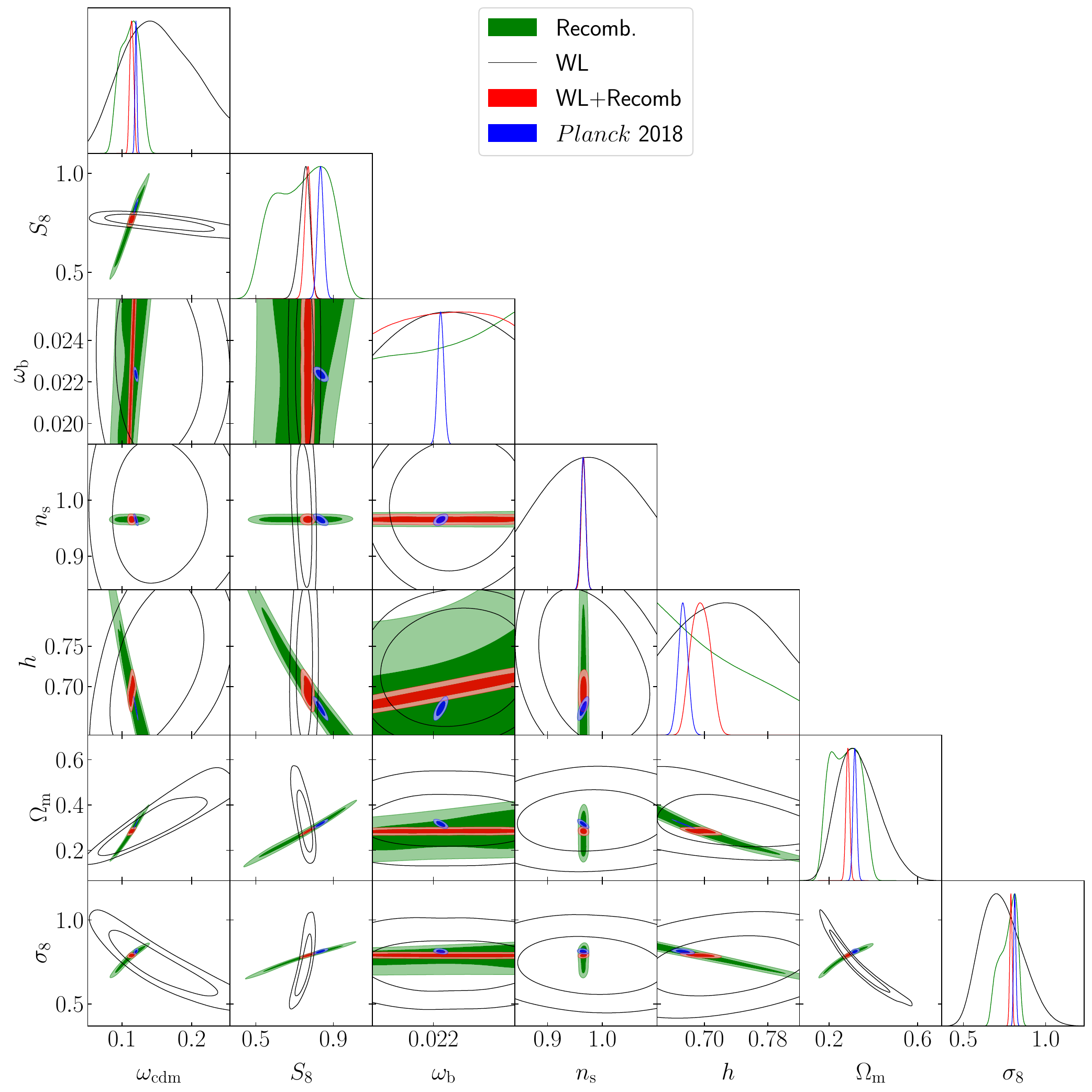} 
        \caption{
        Marginalised posteriors distributions from the analysis of the WL and Recomb data sets. The Recomb data set is composed of data points for $\theta^*$, $A_{\rm s}$, and $n_{\rm s}$ which are inferred via marginalisation of the fiducial \textit{Planck} posteriors \citep{Planck} over the remaining cosmological parameters, as described in Section \ref{Cosmic microwave background anisotropies}. Each panel shows the posterior resulting from a separate analysis of Recomb (green) and weak lensing (black) data sets as well as the combination of both data sets (red). Additionally, we show the \textit{Planck} 2018 contours for reference (blue).}
        \label{fig: CMB}
        \end{figure*}

We present a comparison between the posteriors of a separate traditional analysis of the weak lensing data set and the individual traditional analyses of clustering, Lyman-$\alpha$, and Recomb  data . In Fig. \ref{fig: BOSS} we compare the individual posteriors for weak lensing (black) and clustering (green) with the constraints from the combined analysis of the two data sets (red). Additionally, we show the \textit{Planck} 2018 contours \citep{Planck} for reference (blue). Similarly, we show a comparison between weak lensing and Lyman-$\alpha$ data in Fig. \ref{fig: Lya} and between weak lensing and Recomb data in Fig. \ref{fig: CMB}. We note that the Recomb data set, shown as green contours in Fig. \ref{fig: CMB}, is composed of data points for $\theta^*$, $A_{\rm s}$, and $n_{\rm s}$, which are inferred via marginalisation of the fiducial \textit{Planck} posteriors over the remaining cosmological parameters, as described in Section \ref{Cosmic microwave background anisotropies}.

Our traditional constraints (i.e. with a single set of cosmological parameters governing geometry and growth) are in excellent agreement with the fiducial analyses of the band power spectra in \citet{K1K}, the BOSS DR12 consensus analysis in \citet{BOSS} and the analysis of eBOSS DR14 data in \citet{ebosslkl}. Moreover, the Recomb data analysis recovers the \textit{Planck} posterior for the parameters $n_{\rm{s}}$, $A_{\rm{s}}$ and $\theta^*$, as expected. The $n_{\rm{s}}$ and $A_{\rm{s}}$ posteriors impose  constraints along $S_{\rm{8}}$ or $\sigma_{\rm{8}}$, which are consistent with \textit{Planck} constraints but significantly broader. Similarly, the $\theta^*$ measurement imposes a degenerate constraint in the $h-\omega_{\rm{cdm}}$ plane, which contains the \textit{Planck} 2018 contours (see Fig.~\ref{fig: CMB} for details). 
We observe that for the three combinations of KiDS data with external data sets, i.e 'K1K+Clustering', 'K1K+Ly$\alpha$', and 'K1K+CMB', the posterior contours of the respective combined analysis fall at the intersection of the posteriors of the two individual data sets; see Figs.~\ref{fig: BOSS}, \ref{fig: Lya}, and \ref{fig: CMB}.

\section{Goodness of fit} 
\label{app:Goodness of fit}

\begin{table*}[t] 
\centering
\caption{ Best-fit $\chi^2$ values for all different combinations of data sets considered.}
\label{tab:gof}
\begin{tabular}{llllllll}
 \hline
 \hline
 & \multicolumn{2}{c}{$\chi^2$} &   \multicolumn{2}{c}{DoF} & \multicolumn{2}{c}{$\overline{\chi}^2$} \\ \hline
  
  Data sets & Trad. & Split & Trad. & Split & Trad. & Split \\
  \hline
WL  &   152.34 & 150.76 & 120 - 7 & 120 - 12 & 1.35  & 1.40   \\ 
WL + Clustering  & 157.49 & 156.36 & 130 - 7 & 130 - 12 & 1.28  & 1.33  \\ 
WL + Ly$\alpha$  & 152.37 + 6499.31 & 152.82 + 6499.31 & 6483 - 7 & 6483 - 12 & 1.03 &  1.03   \\ 
WL + Recomb & 151.67 & 152.75 & 123 - 7 & 123 - 12 & 1.32  & 1.37   \\
WL + Clustering + Ly$\alpha$  & 161.33 + 6499.31 & 160.76 + 6499.31  & 6493 - 7 & 6493 - 12 & 1.03  & 1.03 \\
WL + Clustering + Ly$\alpha$ + Recomb & 163.16 + 6499.31 & 161.19 + 6499.31 & 6496 - 7 & 6496 - 12 & 1.03  & 1.03 \\
  \hline
\end{tabular}
\tablefoot{Shown are the best-fit $\chi^2$ values, the corresponding number of degrees of freedom (DoF), and the goodness of fit $\overline{\chi}^2$. Values are given for both the geometry and growth split analysis and the traditional analysis. In the rows reporting on results from Lyman-$\alpha$ forest data we make the contribution to the joint $\chi^2$ from the eBOSS DR14 baseline explicit; see Section \ref{Lyman-a forest and quasars} for more details.}
\end{table*}

In Table \ref{tab:gof} we provide the best-fit $\chi^2$-values, inferred from the maximum of the posterior, for each combination of data sets that we consider, and for both the split analysis of geometry and growth as well as the traditional analysis. Additionally, we show the number of free parameters employed in each of the analyses ($n_{\rm{params}}$), as well as the associated number of data points ($n_{\rm{points}}$). We note that in this work we adopted the fiducial KiDS-1000 cosmology covariance matrix. The fiducial results of \citet{K1K}, however, were obtained with an iterated covariance model which is based on the best-fit parameters of \citet{K1K+BOSS}. While the covariance model has a negligible impact on the posteriors \citep[see][]{K1K}, it does impact the numerical $\chi^2$-value inferred from the maximum of the posterior at the few-percent level. Therefore, the best-fit $\chi^2$-values reported in Table~\ref{tab:gof} are not directly comparable to the ones reported in \citet{K1K}.

It is common practice to assess the goodness of fit under the assumption that the $\chi^2$ statistic follows a $\chi^2$ distribution with the number of degrees of freedom ${\rm DoF} = n_{\rm{points}} - n_{\rm{params}}$. The reduced $\chi^2$, that is defined by $\overline{\chi}^2 = \chi^2/{\rm DoF}$, is then used as a measure of goodness of fit. However, this assumption requires that the data are normally distributed, that the model is linearly dependent on the sampling parameters, and that there is no informative prior on the parameter ranges \citep[see for instance][]{K1K_methodology}. These conditions do not hold in general in cosmological analyses, which require a more sophisticated estimate of the effective number of degrees of freedom which can be inferred for example from mocks or posterior predictive data realisations \citep{chi2, spiegelhalter02, handley19, raveri19, K1K_methodology}.

For the weak lensing data set used in this work, \cite{K1K_methodology} report an effective number of degrees of freedom of $n_{\rm params, eff} = 4.5$. We cannot simply adopt this number since the split of cosmological parameters into two theory regimes and the addition of external data sets is expected to have an impact on the effective number of degrees of freedom. Therefore, we restrict the analysis to an interpretation of the commonly used approximation of the degrees of freedom as the difference between the number of data points and the number of parameters, that is conservative in identifying underfitting models. Nevertheless, we treat the five nuisance parameters that parameterise a shift in the mean of the five KiDS redshift bins as essentially fixed by their strongly informative Gaussian prior and thus we do not count them as free parameters.

The reduced $\chi^2$ for KiDS weak lensing is fairly high. As discussed in detail in \citet{K1K}, this result only occurs for the band power statistic and does not appear to point to a systematically underfitting model. We also applied our split analysis to an earlier KiDS data release \citep[cf.][]{Wright19} and found consistent results with good $\chi^2$. The combination with any of the other data sets acts to reduce the reduced $\chi^2$ closer to unity, such that there is no sign of inconsistency between the probes we combine. 
We observe that the split analyses show a slightly better fit than their traditional counterparts represented by smaller values of $\chi^2$.  However, comparing the reduced $\chi^2$, that takes the increased number of model parameters into account, we do not find a significant preference for the split model. Thus, we conclude that while the split model is better at fitting the data, such improvement is not decisive at justifying the extra added degrees of freedom with respect to the traditional analysis.  

Moreover, we calculate the Bayes factor \citep{bayes_factor},
\begin{equation} \label{eq:bayes_factor}
    \mathcal{B} = \frac{\mathcal{Z_{\rm{trad}}}}{\mathcal{Z_{\rm{split}}}} \, ,
\end{equation}
where $\mathcal{Z}$ is the Bayesian evidence found for each model (see Eq.~\ref{eq:bayes} for reference), 
as well as the difference between the Deviance Information Criterion (DIC; \citealp{gelman04}),
\begin{equation} \label{eq:DIC}
    \rm{DIC} = \frac{1}{2} Var[\chi^2] - \overline{\chi^2} \, ,
\end{equation}
between the traditional and split model; $\Delta {\rm DIC} = {\rm DIC}_{\rm{trad}} - {\rm DIC}_{\rm{split}}$, for each data set. The bar denotes the mean over the sampled $\chi^2$ values. Positive values of $\ln \mathcal{B}$ correspond to more Bayesian evidence in support of the traditional model; negative values to preference for the split model. Models with a lower DIC are preferred by the data since they either possess a lower mean $\chi^2$  (i.e. a better fit of the data) or a lower variance in their $\chi^2$ (i.e. a lower effective number of parameters). Thus, the DIC acts as a combined measurement of goodness of fit and model complexity. Therefore, negative values of  $\Delta{\rm DIC}$ represent a preference for the traditional model while negative values support the split model. 

The obtained values are shown in Table~\ref{tab:model_comp}. In order to asses the significance of the values reported by both metrics we follow the criteria laid out in \citet{Joudaki17}. We interpret $-1 < \ln \mathcal{B} < 1$  and  $\rm{DIC}_{\rm{trad}} - \rm{DIC}_{\rm{split}} < 5$ values as not to be significant enough to prefer either of the models. Therefore, we conclude that neither the Bayes factor nor the DIC show a preference for the split or the traditional model \footnote{This assessment is subject to the nested sampler accuracy in the calculations of the evidence. In order to assure a reliable calculation of the evidence the following settings for \textsc{MultiNest} were used: $\rm{NS\_max\_iter}$: 10000000,  $\rm{NS\_importance\_nested\_sampling}$ : True,  $\rm{NS\_sampling\_efficiency}$ : 0.3,  $\rm{NS\_n\_live\_points}$ : 1000,  $\rm{NS\_evidence\_tolerance}$: 0.1 .}.

\begin{table}[ht] 
\centering
\caption{Model comparison between the traditional and the split model, using the Bayes factor $\mathcal{B}$ and the DIC.}
\label{tab:model_comp}
\begin{tabular}{p{4cm}ll}
 \hline
 \hline
Data set & $\ln \mathcal{B}$ & $\Delta{\rm DIC}$ \\
  \hline
WL  &  0.139 &  0.855\\ 
WL + Clustering  & -0.103 & 1.153  \\ 
WL + Ly$\alpha$  & 0.001 & 1.104\\ 
WL + Recomb &  -0.051 & 1.035 \\
WL + Clustering + Ly$\alpha$  & 0.221 & 1.047 \\
WL + Clustering + Ly$\alpha$ + Recomb & 0.630 &  0.998 \\
  \hline
\end{tabular}
\end{table}

\section{Marginalised one-dimensional  posterior distributions}
\label{app:1D Results}

\begin{figure*} 
\centering
    \includegraphics[width=1\linewidth]{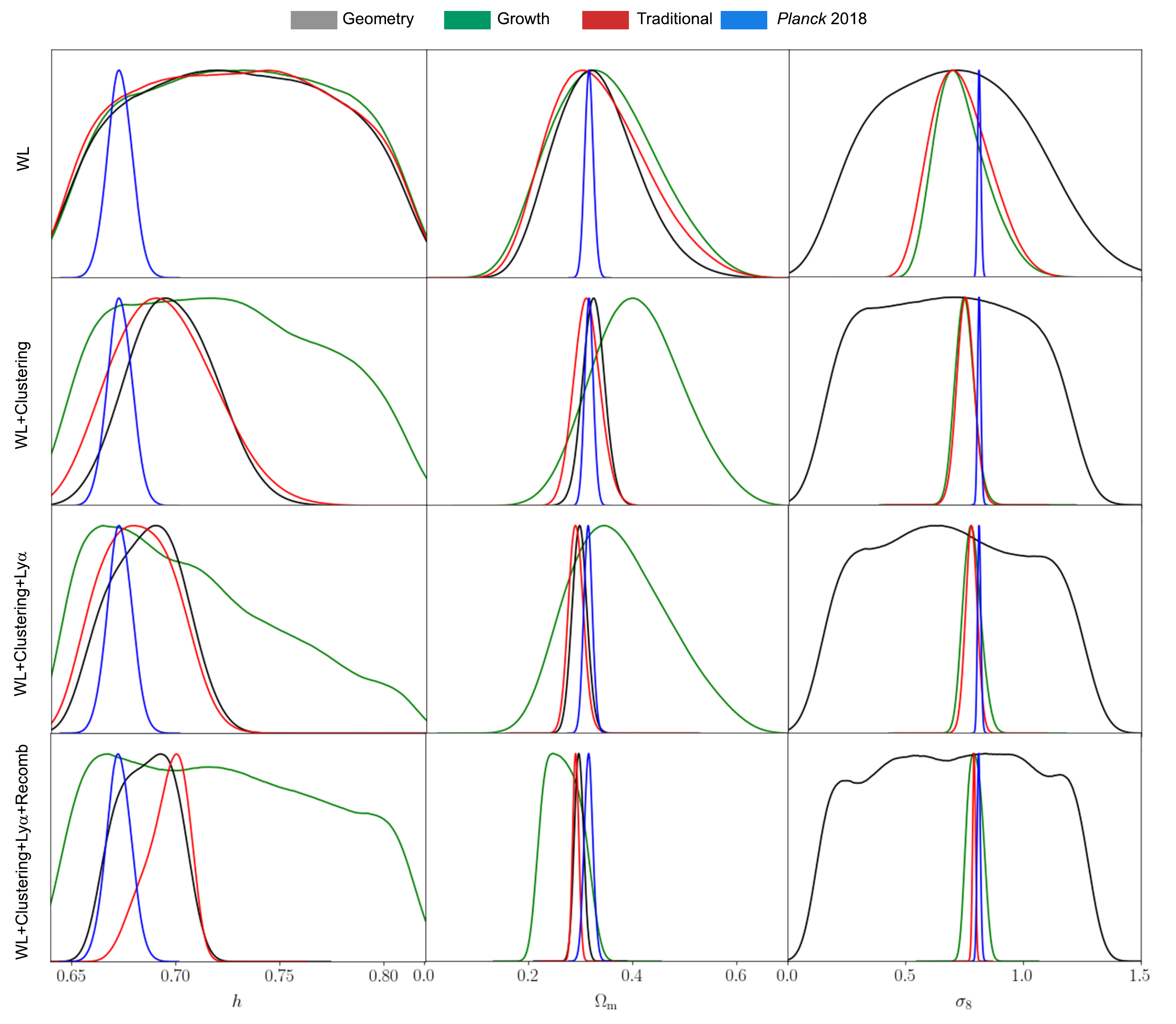} 
    \caption{Marginalised one-dimensional  posterior distributions of the cosmological parameters $h$ (first column), $\Omega_{\rm{m}}$ (second column) and $\sigma_{\rm{8}}$ (third column) for different combinations of data sets (rows). The first row shows the constraints from the WL data set only, while the following rows show the constraints obtained when subsequently adding the Clustering, Ly-$\alpha$, and Recomb data sets.
    In each panel we show a superposition of the constraints from our split analysis of growth and geometry theory regimes (green and grey contours respectively), a traditional analysis with one single set of cosmological parameters (red), and the \textit{Planck} 2018 \citep{Planck} constraints for reference (blue).}
    \label{fig:1D_comp}
\end{figure*}

In Fig.~\ref{fig:1D_comp} we provide the marginalised one-dimensional  posteriors for $h$, $\Omega_{\rm m}$, and $\sigma_8$, corresponding to the posteriors shown in Fig. \ref{fig:2D_comp}, for the combination of weak lensing, clustering, Lyman-$\alpha$, and Recomb data sets. We note that due to how the constraining power is distributed in growth and geometry, growth $h$ and geometry  $\sigma_{\rm{8}}$ constraints are mostly prior-dominated. 

\section{Full cosmological posteriors} 
\label{app:Full Results}

\begin{figure*} 
        \centering
        \includegraphics[width=\linewidth]{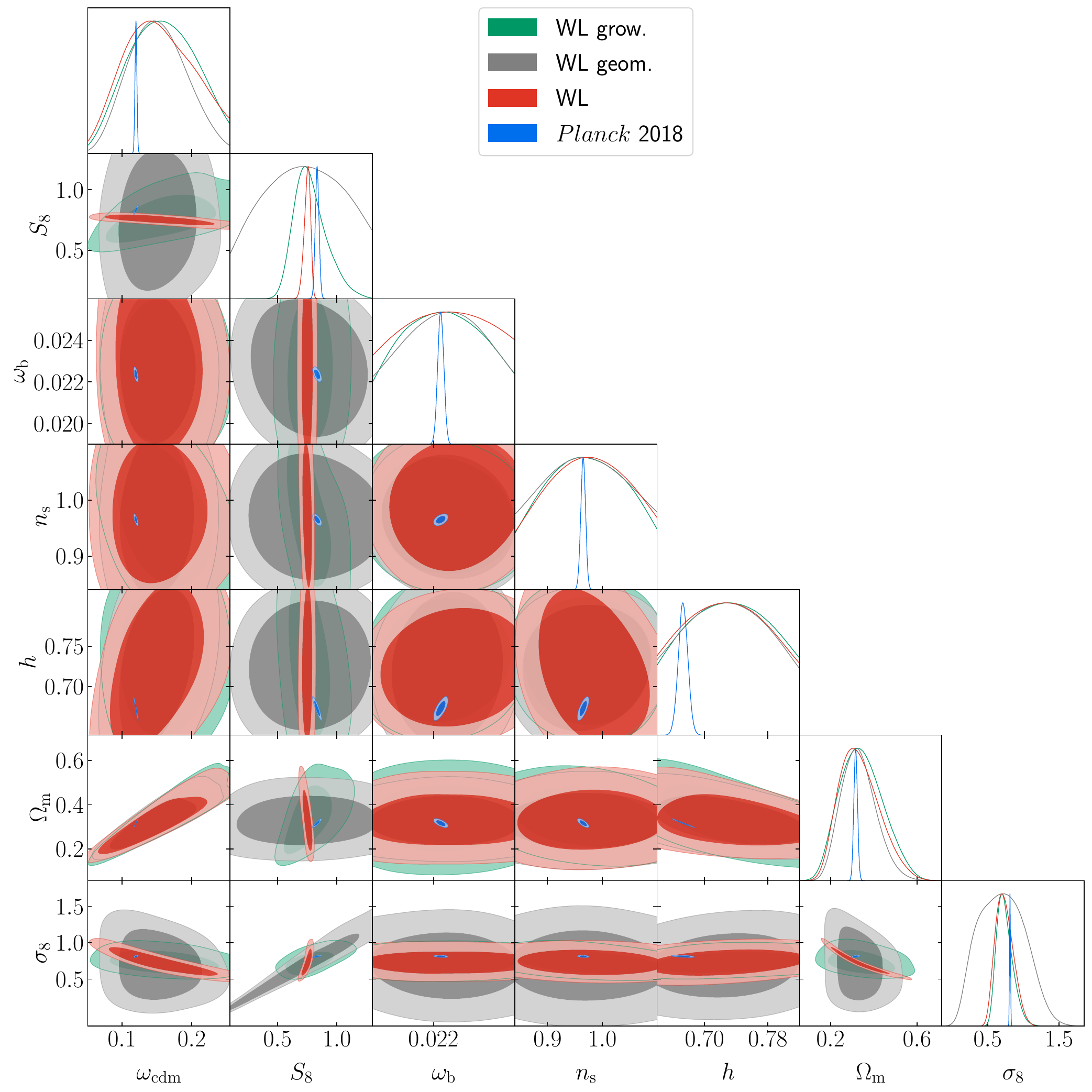} 
        \caption{
        Marginalised posteriors from the analysis of the WL data set for all cosmological parameters. Each panel shows a superposition of the growth (green),  geometry (grey), traditional analysis (red), and the \textit{Planck} 2018 contours \citep{Planck} for reference (blue).}
 \label{fig:WL}
        \end{figure*}

\begin{figure*} 
        \centering
        \includegraphics[width=\linewidth]{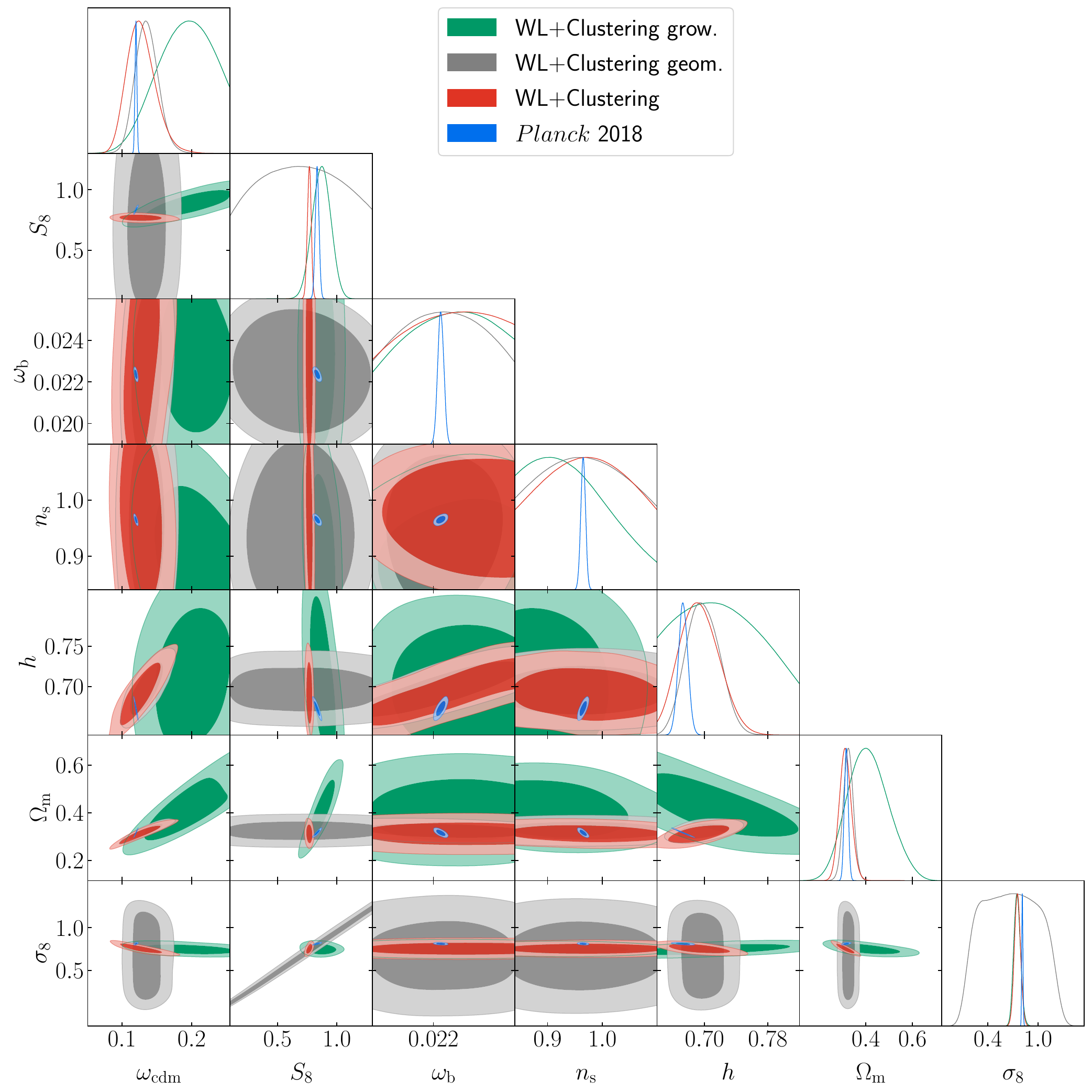}  
        \caption{Marginalised posteriors from the analysis of the WL + Clustering data sets for all cosmological parameters. Each panel shows a superposition of the growth (green),  geometry (grey), traditional analysis (red), and the \textit{Planck} 2018 contours \citep{Planck} for reference (blue).}
 \label{fig:WL+Cl}
        \end{figure*}

\begin{figure*} 
        \centering
        \includegraphics[width=\linewidth]{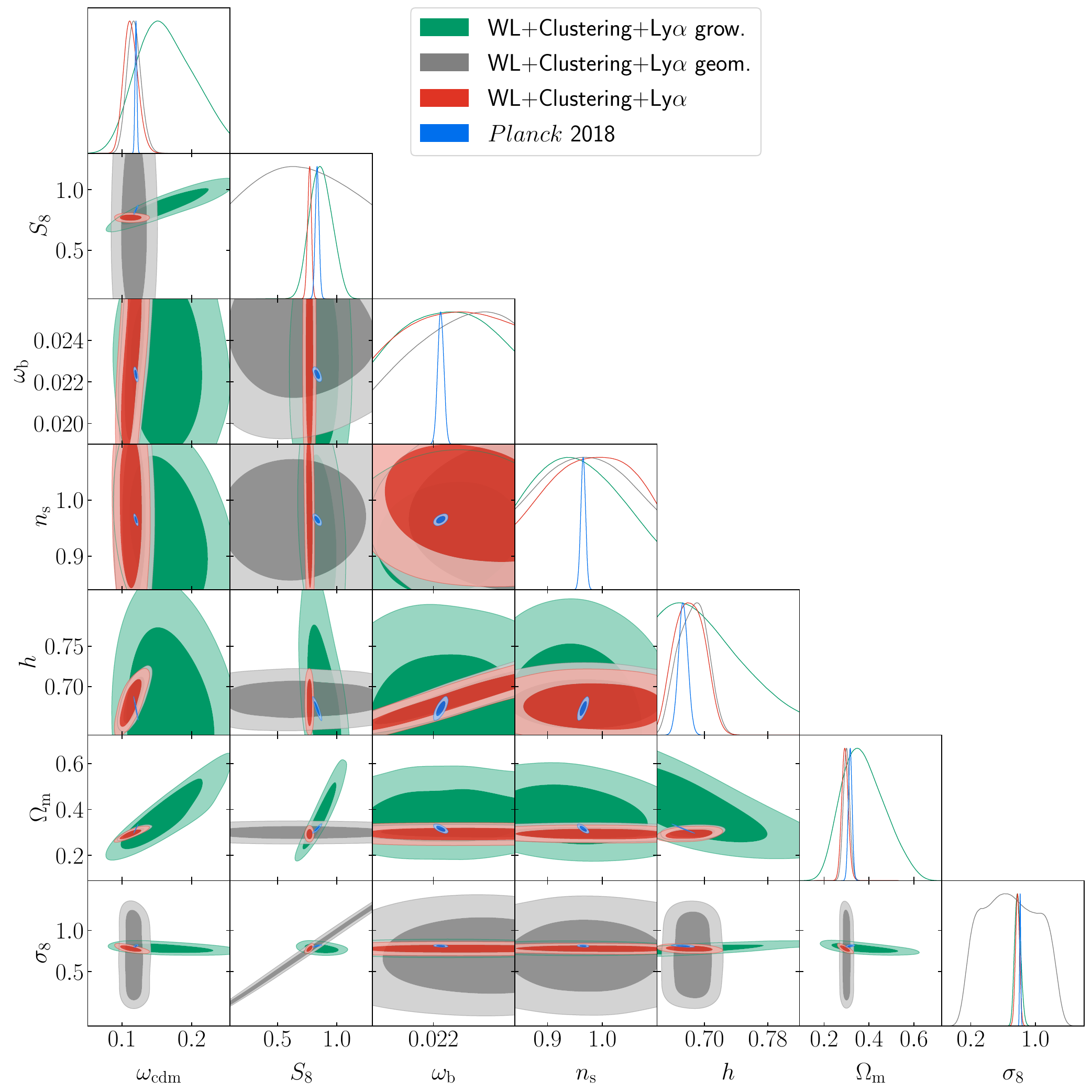}  
        \caption{Marginalised posteriors from the analysis of the WL + Clustering + Ly-$\alpha$ data sets for all cosmological parameters. Each panel shows a superposition of the growth (green),  geometry (grey), traditional analysis (red), and the \textit{Planck} 2018 contours \citep{Planck} for reference (blue).}
 \label{fig:WL+Cl+Ly}
        \end{figure*}

\begin{figure*} 
        \centering
        \includegraphics[width=\linewidth]{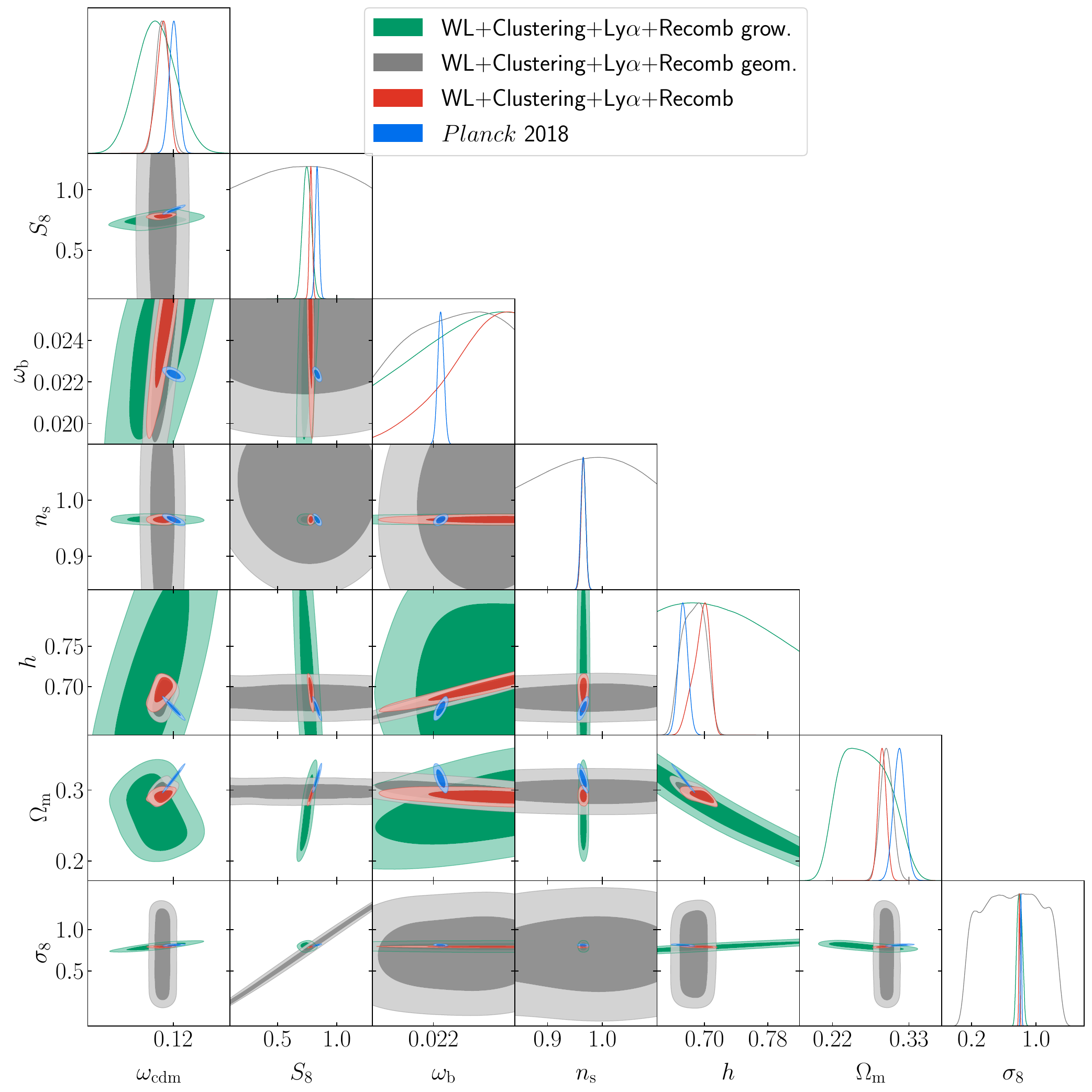}  
        \caption{Marginalised posteriors from the analysis of the WL + Clustering + Recomb data sets for all cosmological parameters. Each panel shows a superposition of the growth (green),  geometry (grey), traditional analysis (red), and the \textit{Planck} 2018 contours for reference (blue).}
 \label{fig:WL+Cl+Ly+CMB}
        \end{figure*}

In this appendix we present the full set of posterior distributions for all cosmological parameters and for each of the studied combinations of data sets shown in Fig. \ref{fig:2D_comp}. In Fig. \ref{fig:WL} we show the marginalised posterior in an analysis of weak lensing data, while Figs. \ref{fig:WL+Cl}, \ref{fig:WL+Cl+Ly}, and \ref{fig:WL+Cl+Ly+CMB} show the constraints in an analysis of weak lensing + clustering, weak lensing + clustering + Lyman-$\alpha$, and weak lensing + clustering + Lyman-$\alpha$ + Recomb data, respectively. In each figure we provide the constraints on geometry and growth parameters from a split analysis of the two theory regimes (grey and green contours), as well the traditional analysis with one set of cosmological parameters (red). Additionally, we show the \textit{Planck} 2018 contours \citep{Planck} for reference (blue).

\end{document}